\documentclass[12pt]{article}
\usepackage[top=3cm, bottom=3cm, left=2cm, right=2cm]{geometry}
\usepackage[usenames,dvipsnames,svgnames]{xcolor}
\usepackage[numbers,sort&compress]{natbib}

\pdfoutput=1
\usepackage{amsmath,amssymb}
\usepackage{graphicx}
\usepackage{subfig}
\usepackage[utf8]{inputenc}
\usepackage{cancel}
\usepackage[colorlinks=true]{hyperref}
\usepackage{color}
\usepackage{comment}
\usepackage{array,multirow}
\usepackage{float}
\usepackage[normalem]{ulem}

\title{Reopening the $Z$ portal with semi-annihilations}

\author{
María J. Dom\'inguez$^{1,}$\footnote{\href{mailto:mjose.dominguez@udea.edu.co}{mjose.dominguez@udea.edu.co}},\,
Oscar Rodr\'iguez$^{1,2,}$\footnote{\href{mailto:oscara.rodriguez@udea.edu.co}{oscara.rodriguez@udea.edu.co}},\, 
\'Oscar Zapata$^{1,}$\footnote{\href{mailto:oalberto.zapata@udea.edu.co}{oalberto.zapata@udea.edu.co}}\\
\textit{\small $^1$Instituto de F\'isica, Universidad de Antioquia,}\\ 
\textit{\small Calle 70 \# 52-21, A.A. 1226, Medell\'in, Colombia}\\
\textit{\small $^2$Facultad de Ingenier\'ias, Universidad de San Buenaventura,}\\ 
\textit{\small Carrera 56C \# 51-110, Medell\'in, Colombia}
}
\date{\today}
\begin{document}
\maketitle

\begin{abstract}
In one-component dark matter (DM) scenarios is commonly assumed that a scalar WIMP must either be part of an $SU(2)_L$ multiplet with zero hypercharge or have suppressed vector interactions with the $Z$ gauge boson to circumvent stringent direct detection (DD) bounds. In this work, we demonstrate that multi-component scenarios with a dark scalar doublet exhibiting vector-like interactions with the $Z$ boson are also compatible with bounds arising from DD searches. Specifically, we consider a simple extension of the Standard Model wherein the dark sector comprises a doublet and a complex singlet $\phi$, both charged under a $Z_6$ symmetry. We find that semi-annihilation processes drastically reduce the relic abundance of the neutral component of the doublet, $H^0$, sufficiently attenuating the effects of its large $Z$-mediated elastic scattering cross-section with nucleons to satisfy the DD constraints. Although the contribution of $H^0$ to the total relic abundance is nearly negligible, with $\phi$ dominating, both dark matter components are expected to be detectable in ongoing and future DD experiments.  
The viability of the model is tested against several theoretical and experimental constraints, resulting in a parameter space featuring a non-degenerate mass spectrum at the electroweak scale.
\end{abstract}

\section{Introduction}
The observed relic abundance of dark matter (DM)~\cite{Planck:2018vyg} finds a simple and appealing explanation through a new stable particle characterized by electroweak-scale mass and interactions~\cite{Steigman:1984ac}. This particle achieves chemical equilibrium with the Standard Model (SM) in the early universe, following the WIMP mechanism~\cite{Jungman:1995df, Bertone:2004pz}. In this context, the SM Higgs and $Z$ bosons may act as mediators between the dark and visible sectors, giving rise to the Higgs-portal~\cite{Patt:2006fw,Arcadi:2019lka} and $Z$-portal~\cite{Arcadi:2014lta} models.
Among the simplest models accommodating the Higgs portal are those that introduce a SM singlet, either a scalar~\cite{Silveira:1985rk,McDonald:1993ex,Burgess:2000yq} or a fermion~\cite{Kim:2006af,Kim:2008pp,Baek:2011aa}. Conversely, the archetype of a purely $Z$-mediated DM model is the neutralino~\cite{Ellis:1983ew, Jungman:1995df}. Notably, renormalizable DM models encompassing both Higgs and $Z$ portals, such as the inert doublet model (IDM)~\cite{Deshpande:1977rw,Barbieri:2006dq}, can be easily found. 

Direct searches~\cite{Schumann:2019eaa} have been crucial in probing large portions of the parameter space of WIMP models. Indeed, in models having an open $Z$-portal they exclude DM candidates that can elastically scatter on nuclei via a tree-level $Z$ boson exchange due to a spin-independent (SI) cross section lying orders of magnitude above the current bounds~\cite{PandaX-4T:2021bab, LZ:2022lsv, XENON:2023cxc}. 
This situation materializes, for instance, in the IDM where the neutral component of the inert doublet acts as the DM candidate,  leading to a large scattering cross section with nuclei of the order of $10^{-39}$ cm$^2$~\cite{Barbieri:2006dq}. Nonetheless, it is only when the CP components of the doublet are non-degenerate (leading to a non-diagonal coupling to $Z$ boson) that the stringent direct detection (DD) limits can be circumvented. This enables the explanation of the observed DM abundance for masses near the Higgs resonance and exceeding 500 GeV~\cite{LopezHonorez:2006gr}.

DM in the Universe may not necessarily made up of a single particle and instead populated by several species accounting for the total abundance~\cite{Boehm:2003ha,Ma:2006uv,Cao:2007fy,Hur:2007ur,Lee:2008pc,Zurek:2008qg,Barger:2008jx,Profumo:2009tb}. In these frameworks, new DM processes~\cite{Hambye:2008bq,DEramo:2010keq} such as conversions and semi-annihilations typically arise, modifying not only the production of each component but also their interactions with the visible sector, thus providing a reason for the lack of DD signatures in single DM component scenarios. 
Certainly, recent phenomenological studies on multi-component WIMP models (see~{\it e.g.} Refs.  \cite{Bernal:2018aon,Poulin:2018kap,Borah:2019aeq,Nanda:2019nqy,Bhattacharya:2018cgx, Bhattacharya:2019fgs, Betancur:2020fdl,Hernandez-Sanchez:2020aop, Belanger:2020hyh, Choi:2021yps, Belanger:2021lwd, DiazSaez:2021pmg, Mohamadnejad:2021tke, Chakrabarty:2021kmr, Ho:2022erb, Bhattacharya:2022wtr, DuttaBanik:2020jrj, Hernandez-Sanchez:2022dnn, BasiBeneito:2020mdr, Hall:2019rld, Hall:2021zsk, Yaguna:2021vhb, Yaguna:2021rds, Das:2022oyx, Belanger:2022esk, Hosseini:2023qwu, Yaguna:2023kyu}) have shown that their compatibility with current experimental data is possible even with DM masses significantly less than 1 TeV. 

In this work, we analyze a two-component DM model with two scalar candidates: a SM singlet $\phi$ and the neutral component $H^0$ of a second weak-isospin doublet $H_2$, both charged under a $Z_6$ symmetry\footnote{For scenarios involving a $Z_6$ symmetry responsible for DM stability, see Refs.~\cite{Yaguna:2019cvp,VanDong:2020bkg, Yaguna:2021vhb, Yaguna:2024jor}.}. Both candidates retain their complex nature even after electroweak (EW) symmetry breaking, implying that $H^0$ continues having diagonal gauge interactions with the $Z$ boson. We demonstrate that the semi-annihilation processes induced by the interaction term $H^\dagger_2 H_1 \phi^2$ cause a large suppression on the $H^0$ abundance in such a way the expected number of events associated with $H^0$ in DD experiments, such as LUX-ZEPLIN (LZ)~\cite{LZ:2022lsv}, can lie below the current upper bound. Moreover, this conclusion is obtained guaranteeing that all constraints imposed on the model are fulfilled in a range of non-degenerate DM masses around the EW scale. In this way, the doublet candidate not only emerges as a valid candidate but also can leave signatures in current and future DM experiments. 

The rest of the paper is organized as follows. In the next section, we present the model and study the effect of the interactions allowed by the $Z_6$ symmetry on the DM relic densities, as well as the DD of the DM candidates. Special emphasis is placed on calculating the expected number of events in Xenon-based experiments. 
In section~\ref{sec:Numerics}, we describe the theoretical and experimental constraints that must be satisfied and determine the viable parameter space through random scans. 
This analysis also includes an investigation of detection prospects.
Finally,  we present our conclusions in section~\ref{sec:Conclusions}.

\section{The model}\label{sec:model}
The model we consider enlarges the SM by introducing a dark scalar sector made up of a second Higgs doublet $H_2$ and a complex gauge singlet $\phi$. Besides, an exact $Z_6$ symmetry is introduced such that
the dark sector fields are charged under $Z_6$ whereas the SM fields transform trivially. In order to have two DM candidates, besides imposing that the new scalar fields do not acquire a non-zero vacuum expectation value, the neutral component of $H_2$ and $\phi$ must not mix with each other\footnote{Notice that the model presented in Ref.~\cite{Belanger:2021lwd} considers the neutral components of $H^0$ to be non-degenerate.}.  The charge assignment assuring these conditions becomes\footnote{The discrete symmetry can be promoted to a $U(1)_X$ global symmetry, leading to the charge assignment $X(H_2)=2X(\phi)$ and $X({\rm SM})=0$, such that $2X(\phi)-X(H_2)=0$. Recall that this possibility opens the door to considering a dark asymmetry~\cite{Graesser:2011wi}  within a two-component scenario~\cite{BasiBeneito:2022qxd, Herrero-Garcia:2023lhv}.}
\begin{align}
Z_6(\phi)=\omega_6,\ \ Z_6(H_2)=\omega_6^2,\ \ \ \ \omega_6=e^{i\pi/3}. \label{eq:chargeAssignment}
\end{align}
Consequently, the most general $Z_6$-invariant scalar potential reads
\begin{align}
\mathcal{V}&=\, -\mu^2_1|H_1|^2+\lambda_1|H_1|^4 + \mu^2_2|H_2|^2 + \lambda_2|H_2|^4 + \mu_{\phi}^2|\phi|^2+\lambda_{\phi}|\phi|^4 + \lambda_3 |H_1|^2|H_2|^2\nonumber\\
  & \, + \lambda_4 |H_1^\dagger H_2|^2 + \lambda_6 |H_2|^2|\phi|^2+\frac{1}{2}\lambda_7\left( H_2^\dagger H_1\phi^2+\mathrm{h.c.}\right) +\lambda_8 |H_1|^2|\phi|^2,\label{eq:Z6lag}
\end{align} 
where $H_1$ represents the SM Higgs doublet and $\lambda_7$ is real without loss of generality, achievable through field redefinitions of $\phi$, $H_1$, or $H_2$. 
Working in the unitary gauge,
\begin{align}
H_1=\begin{bmatrix}
0\\
\frac{1}{\sqrt{2}}(v+h)
\end{bmatrix},\ \
H_2=\begin{pmatrix}
H^+\\
H^0
\end{pmatrix},
\end{align}
with $v=246$ GeV, the scalar spectrum is given by
\begin{align}\label{eq:Mscalars}
m_{h}^{2} &=  2\lambda_1v^2,\\
m_{\phi}^{2} &=  \mu_{\phi}^{2} +\dfrac{v^{2}}{2}\lambda_8 ,\\
m_{H^\pm}^{2} &=  \mu_{2}^{2} +\dfrac{v^{2}}{2}\lambda_3 ,\\
m_{H^0}^{2} &= \mu_{2}^{2} + v^{2}\lambda_L,
\end{align}
with $\lambda_L\equiv (\lambda_3+\lambda_4)/2$. When $\phi$ is the lightest particle we impose the kinematic relation $m_{H^0}<2m_\phi$ to ensure the stability of $H^0$.  
Notice that the absence of the term $[\lambda_5(H_1^\dagger H_2)^2+\text{h.c.}]$ entails that the $H^0$ field remains complex, {\it i.e.} the CP even and odd components have the same mass.
Fixing the Higgs boson mass $m_h$ at 125 GeV~\cite{ATLAS:2015yey}, the model includes 9 new free parameters:  6 dimensionless $(\lambda_2,\lambda_3,\lambda_6,\lambda_8,\lambda_\phi,\lambda_L)$ and 3 dimensionful chosen to be the dark scalar masses. 

\begin{figure}[t]
\centering
\includegraphics[scale=0.9]{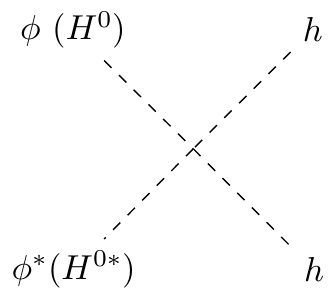}\hspace{0.4cm}
\includegraphics[scale=0.9]{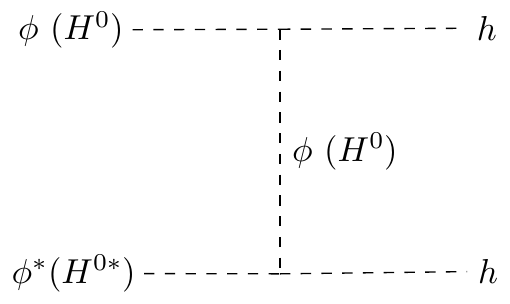}\hspace{0.4cm}
\includegraphics[scale=0.9]{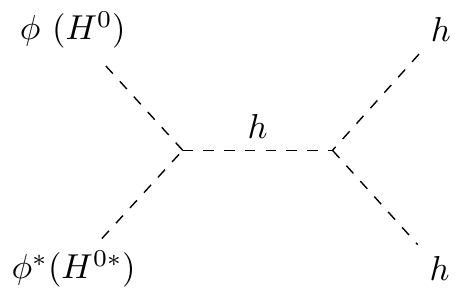}\\ 
\includegraphics[scale=0.9]{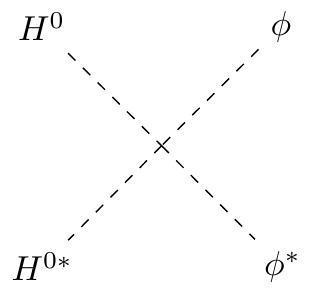}\hspace{0.4cm}
\includegraphics[scale=0.9]{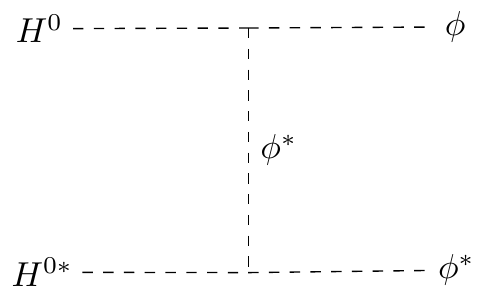}\hspace{0.4cm}
\includegraphics[scale=0.9]{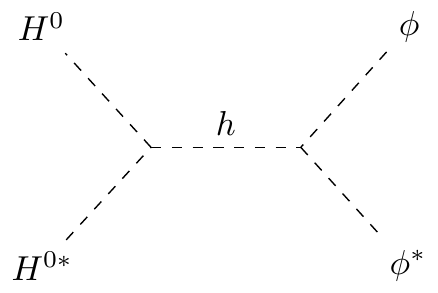}
\caption{Top: DM self-annihilation processes mediated by the Higgs portal interactions. Final states as $WW,ZZ,hZ,f\bar{f}$ can be also present in the $s$-channel processes. Bottom: DM conversion processes mediated by the $\lambda_6$ (left panel), $\lambda_7$ (middle panel) and the Higgs portal (right panel) interactions. }
\label{fig:HP-anni-conv}
\end{figure}

\begin{figure}[t]
\centering
\includegraphics[scale=0.9]{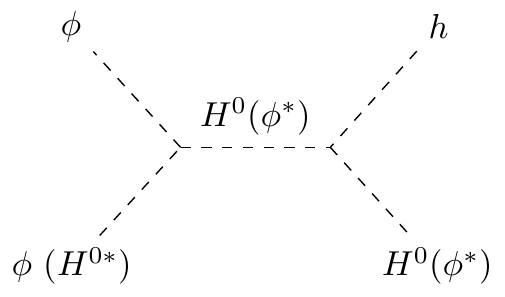}\vspace{0.4cm}
\includegraphics[scale=0.9]{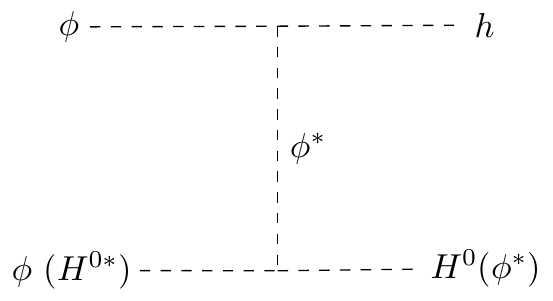}\hspace{0.4cm}
\includegraphics[scale=0.9]{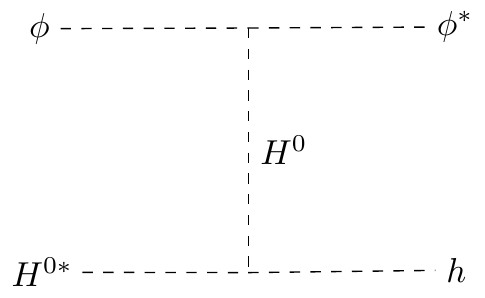}\\
\includegraphics[scale=0.9]{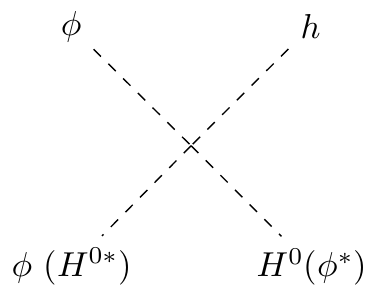}\hspace{0.4cm}
\includegraphics[scale=0.95]{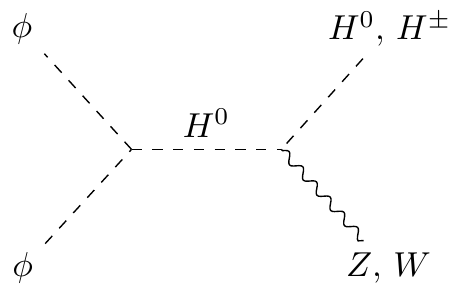}\hspace{0.4cm}
\includegraphics[scale=0.95]{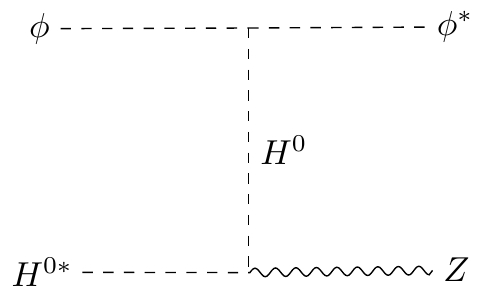}
\caption{DM semi-annihilation processes mediated by the interplay of Higgs portal and $\lambda_7$ interactions (top panels), by only the $\lambda_7$ interaction (bottom left panel) and by the interplay between the $\lambda_7$ and the gauge portal interactions (bottom middle and right panels).}
\label{fig:semi}
\end{figure}

The set of new scalar interactions induces annihilation, semi-annihilation and conversion processes which affect DM relic densities in distinctive ways.  First, the usual Higgs portal interactions, $(\lambda_3,\lambda_4)$ for the doublet and $\lambda_8$ for the singlet, couple the DM particles with the SM ones, leading to DM self-annihilation processes (see Fig.~\ref{fig:HP-anni-conv}).  
Secondly, as is shown in Fig.~\ref{fig:HP-anni-conv}, DM conversion processes can arise in three different ways: from the quartic terms $\lambda_6$ and $\lambda_7$, and from the interplay of the two Higgs portal interaction terms. Thirdly, the Higgs portal interactions also play a role in the semi-annihilation processes when combined with the $\lambda_7$ interaction, although such processes appear independently of the Higgs portal (see Fig.~\ref{fig:semi})\footnote{See Ref.~\cite{Beauchesne:2024vbo} for a study of semi-annihilations in models including one or two scalar multiplets in the dark sector.}. 
Finally, the self-interacting terms, $\lambda_\phi$ and $\lambda_2$, although not relevant for the DM phenomenology, they play a key role in ensuring the theoretical consistency of the model.

On the other hand, gauge interactions of the doublet $H_2$ not only cause (co)annihilation processes but also semi-annihilation processes. The first set involves new processes such as $H^0H^0\to WW$ but also as those mediated by the Higgs portal $H^0H^0\to hh$, among others.  The second set brings with the purely new type of processes involving a gauge boson in the final state thanks to the presence of the $\lambda_7$ interaction term, as displayed in the bottom panels (center and right) of Fig.~\ref{fig:semi}.    
As will be shown in the next section, this type of processes is fundamental to evade the DD constraints. 

Table I summarizes all the $2\rightarrow 2$ processes that may affect the relic densities of $\phi$ and $H^0$. These processes are classified according to the fields involved and to the number of SM particles present in the final state. To this end, $\phi$/$\phi^*$ and $H_2$/$H_2^*$ are assumed to belong, respectively, to sectors 1 and 2, whereas the SM particles belong to sector 0. In this way, processes of the type $1100$, $2200$ and $1200$ correspond to (co)annihilations, whereas $1110$, $2220$, $1120$, $2210$, $1210$ and $1220$ denote semi-annihilation processes. DM conversion processes are of the type $1122$ and $2211$. 

\begin{table}[t]
\begin{center}
        \begin{tabular}{c |c}
      $\phi$ processes   & Type  \\
      \hline
        $\phi+\phi^*\to SM + SM$ & $1100$\\   
        $\phi+\phi^* \to H^0+H^{0*}$  & $1122$\\
        $\phi+\phi\to H^0 + h(Z),H^\pm + W^\mp$ & $1120$\\
    \end{tabular}\hspace{2cm}
    \begin{tabular}{c |c}
      $H^0$ processes   & Type  \\
      \hline 
        $H^0+H^{0*}\to SM + SM$ & $2200$\\
        $H^0+H^{0*} \to \phi+\phi^*$  & $2211$\\
        $H^0 + h \to \phi+\phi$ & $2011$\\
        $H^{0*}+\phi \to \phi^*+h(Z)$  & $1210$\\
    \end{tabular}
    \caption{The 2$\to$ 2 processes allowed (at tree-level) by the $Z_6$ symmetry and that can modify the relic density of $\phi$ (left) and $H^0$ (right). $Z$ and $W^\pm$ denote the EW gauge bosons whereas $h$ stands for the SM Higgs boson. Conjugate and inverse processes are not shown.}
    \label{tab:processesZ4}
\end{center}
\end{table}

\subsection{Relic abundance}
The Boltzmann equations for the model can then be written down as
\begin{align}
\frac{dn_\phi}{dt}&=-\sigma^{1100}_v\left(n^2_\phi-\bar{n}^2_\phi\right)-\sigma^{1120}_v\left(n^2_\phi-n_{H^0}\frac{\bar{n}^2_\phi}{\bar{n}_{H^0}}\right)-\sigma^{1122}_v\left(n^2_\phi-n^2_{H^0}\frac{\bar{n}^2_\phi}{\bar{n}^2_{H^0}}\right)-3Hn_\phi,\\
\frac{dn_{H^0}}{dt}&=-\sigma^{2200}_v\left(n^2_{H^0}-\bar{n}^2_{H^0}\right)-\sigma^{2211}_v\left(n^2_{H^0}-n_\phi\frac{\bar{n}^2_{H^0}}{\bar{n}^2_\phi}\right)-\frac{1}{2}\sigma^{1210}_v\left(n_\phi n_{H^0}-n_\phi \bar{n}_{H^0}\right)\notag\\
&-\sigma^{1120}_v\left(n^2_\phi-n_{H^0}\frac{\bar{n}^2_\phi}{\bar{n}_{H^0}}\right)-3Hn_{H^0}, \label{eq:boltzmannEqs}
\end{align}
where $\sigma_v^{abcd}$ stands for the thermally averaged cross section, which satisfies
$\bar{n}_a\bar{n}_b\sigma^{abcd}_v=\bar{n}_c\bar{n}_d\sigma^{cdab}_v$, 
$H$ is the Hubble constant, $n_{\psi}$ denotes the number density of the $\psi$ field and $\bar{n}_{\psi}$ its the respective equilibrium value. To numerically solve these equations and obtain the relic densities of $\phi$ and $H^0$ (denoted here as $\Omega_\phi$ and $\Omega_{H^0}$, respectively), we used {\tt MicrOMEGAs}~\cite{Belanger:2014vza, Alguero:2023zol} (via {\tt LanHEP}~\cite{Semenov:2014rea}) which automatically takes into account all the relevant processes for a given model. 

\begin{figure}[t]
\centering
\includegraphics[scale=0.45]{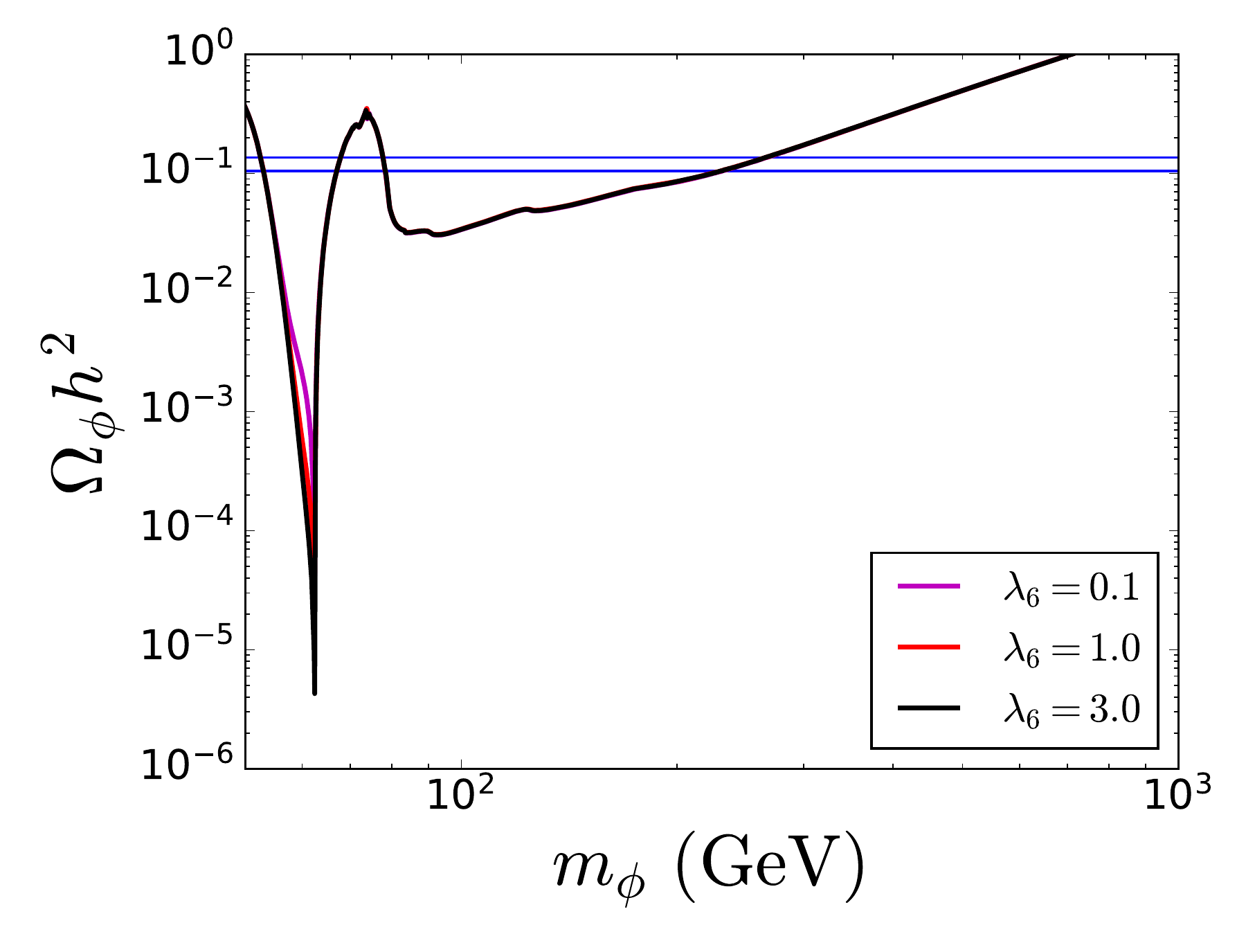}
\includegraphics[scale=0.45]{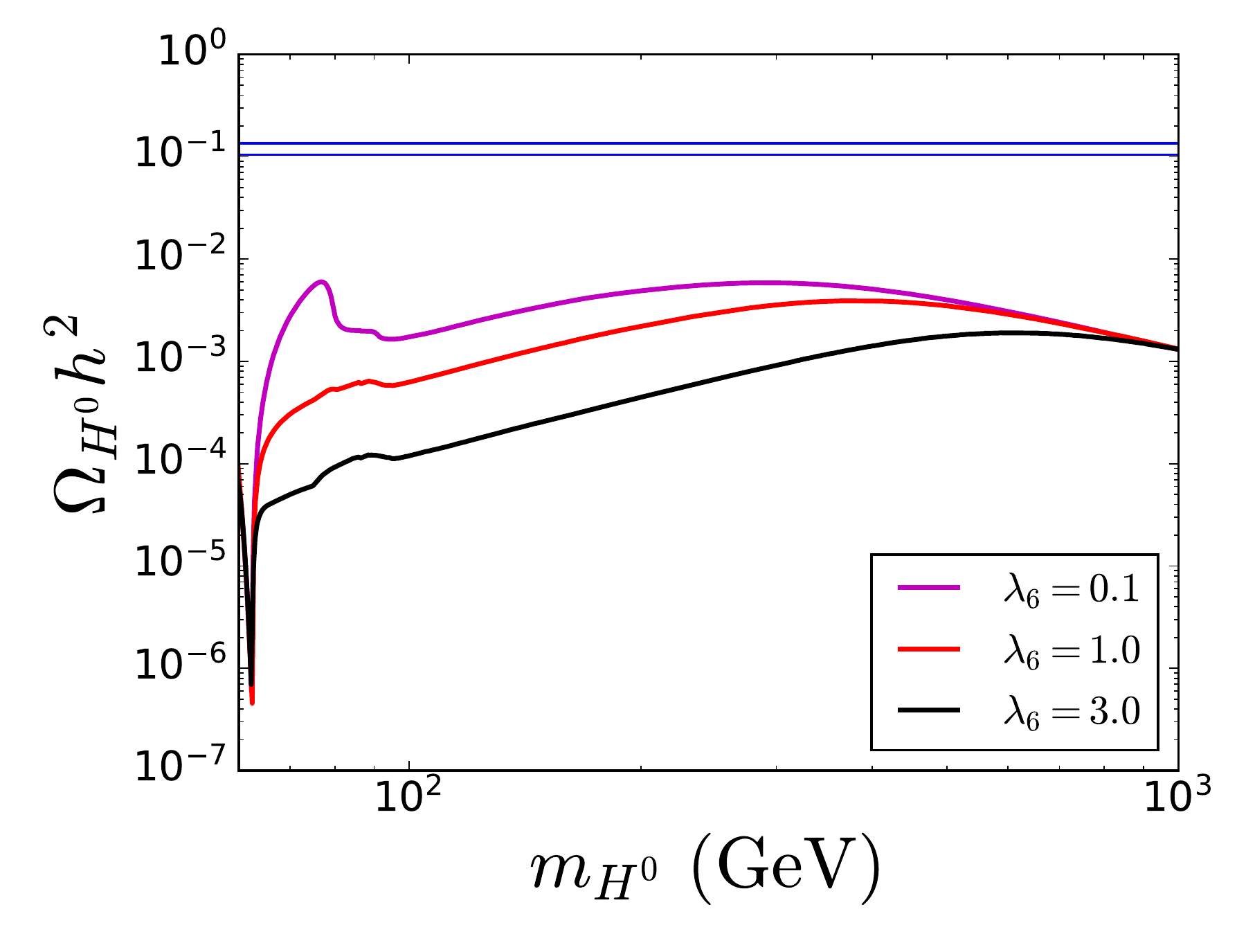}\\
\includegraphics[scale=0.45]{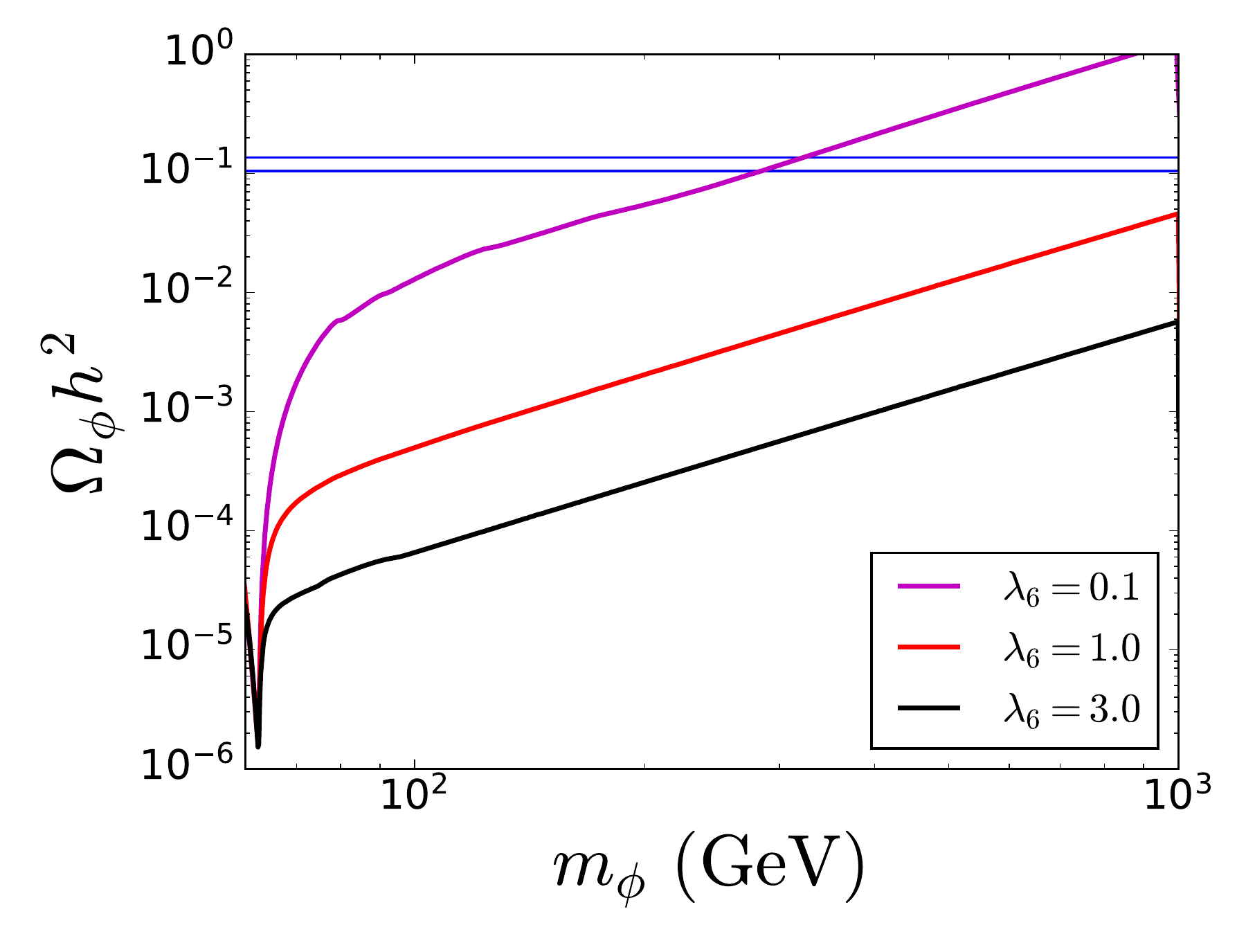}
\includegraphics[scale=0.45]{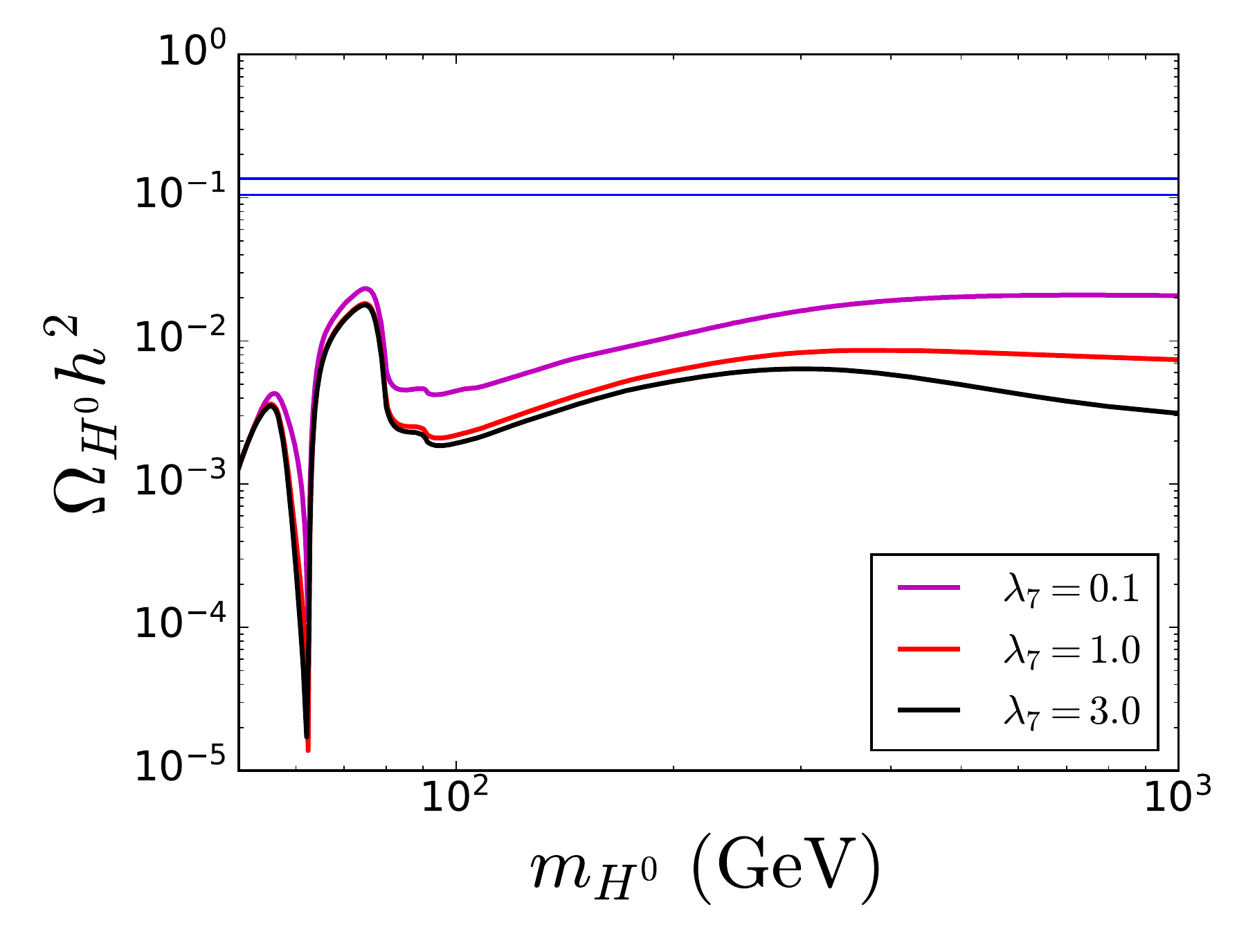}
\caption{Effect of $\lambda_6$-mediated DM conversion on $\Omega_{\phi}$ and $\Omega_{H^0}$ for $m_{H^0}/m_\phi= 1.2$ (top panels) and  $m_\phi/m_{H^0}= 1.2$ (bottom panels). In each plot $m_{H^\pm}/m_{H^0}=1.1$, $\lambda_7=0$ and $\lambda_8=\lambda_L=0.1$ have been set.}
\label{fig:relicdensity1}
\end{figure}

We begin our analysis by considering a set of reference models to understand how the new conversion and semi-annihilation processes modify the DM abundances during freeze-out. In what follows, we fix the $\lambda_8$ and $\lambda_L$ scalar couplings to $0.1$ and $m_{H^\pm}/m_{H^0}=1.1$. Two mass hierarchies are considered: $m_\phi<m_{H^0}<m_{H^\pm}$ and  $m_{H^0}<m_{H^\pm}<m_\phi$. To observe the effect on $\Omega_\phi$ and $\Omega_{H^0}$ due to the DM conversion induced by $\lambda_6$, we take $\lambda_7=0$. On the other hand, fixing $\lambda_6=0$ but keeping $\lambda_7\neq 0$, we are able to appreciate the effect of the DM semi-annihilation processes. 

\begin{figure}[t]
\centering
\includegraphics[scale=0.45]{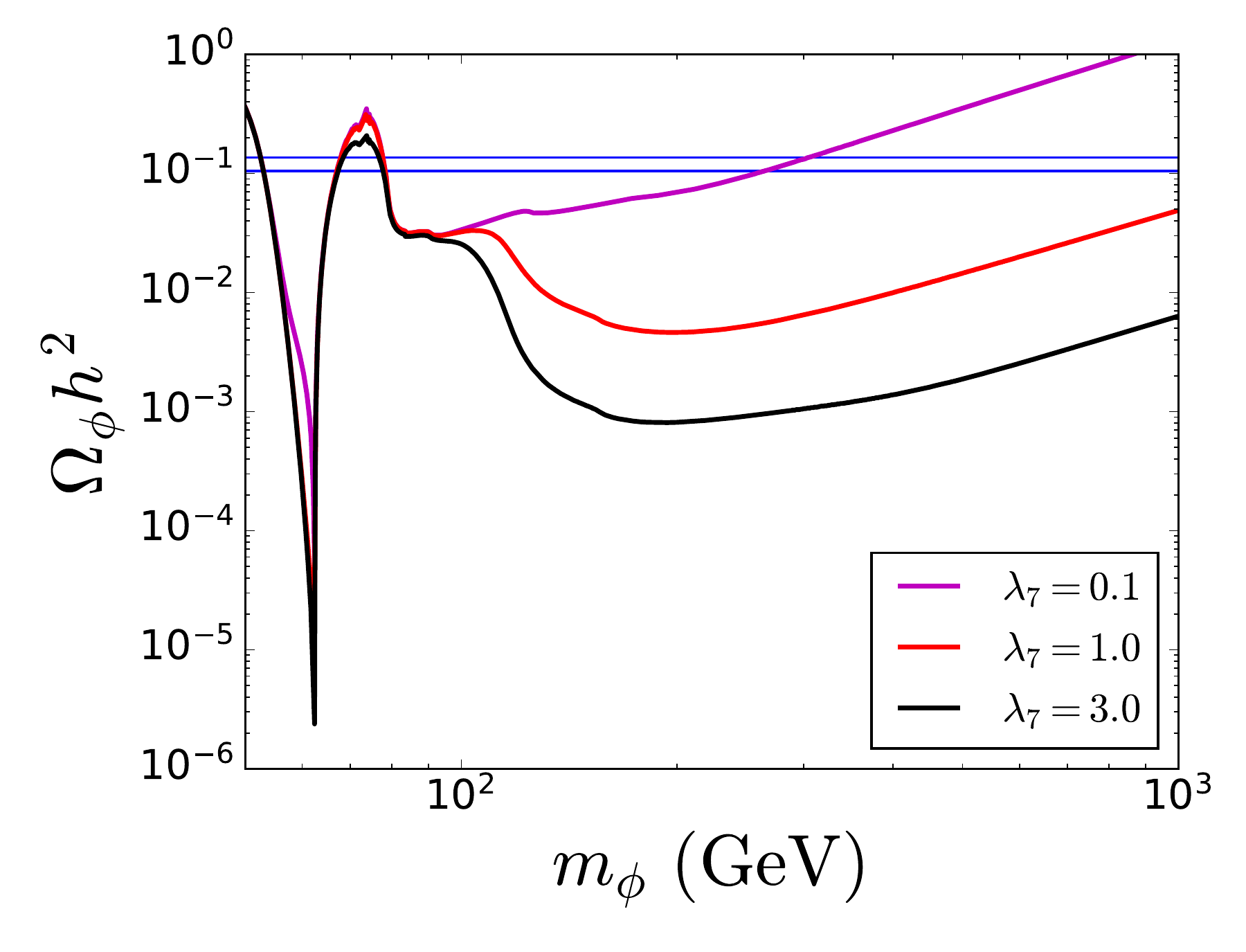}
\includegraphics[scale=0.45]{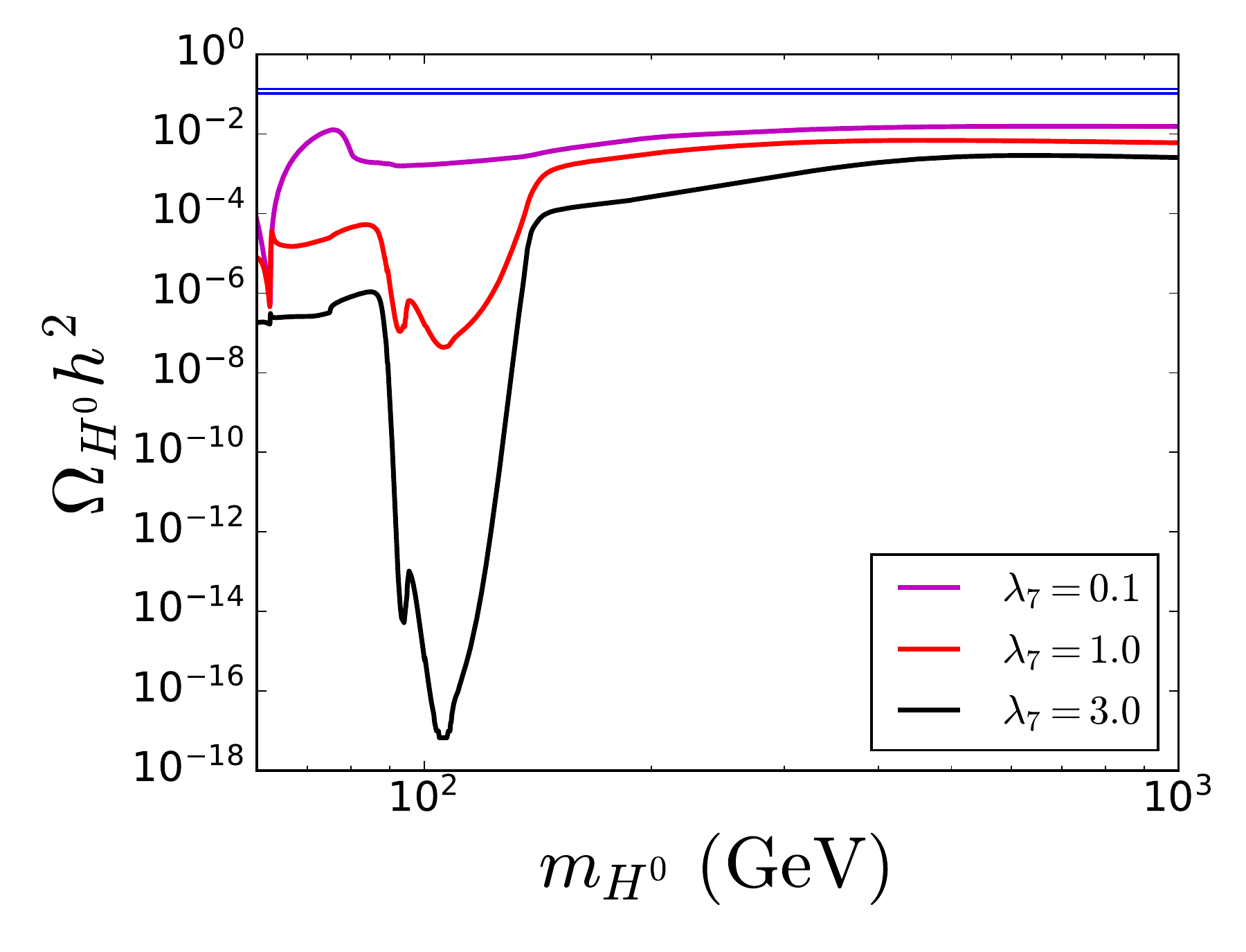}\\
\includegraphics[scale=0.45]{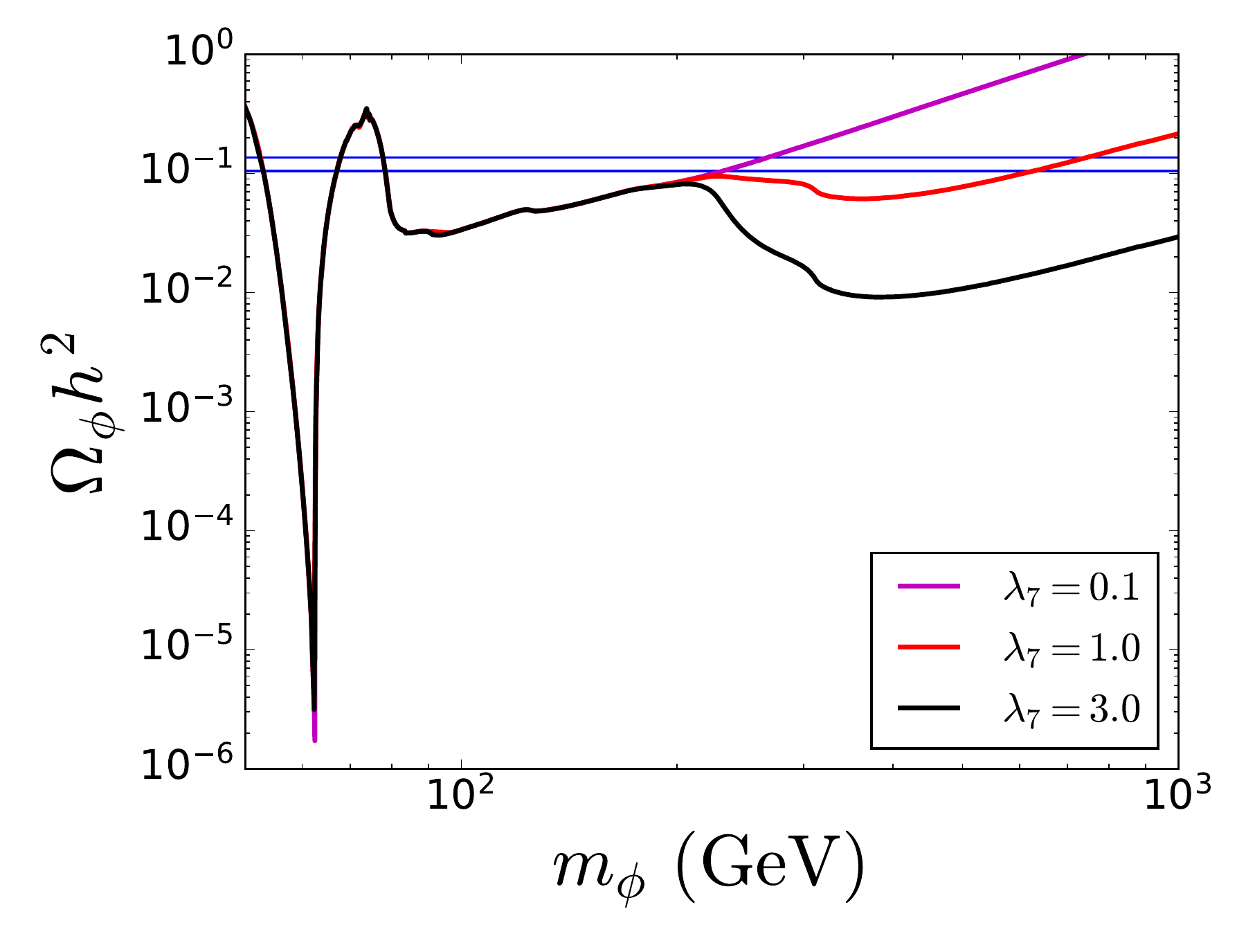}
\includegraphics[scale=0.45]{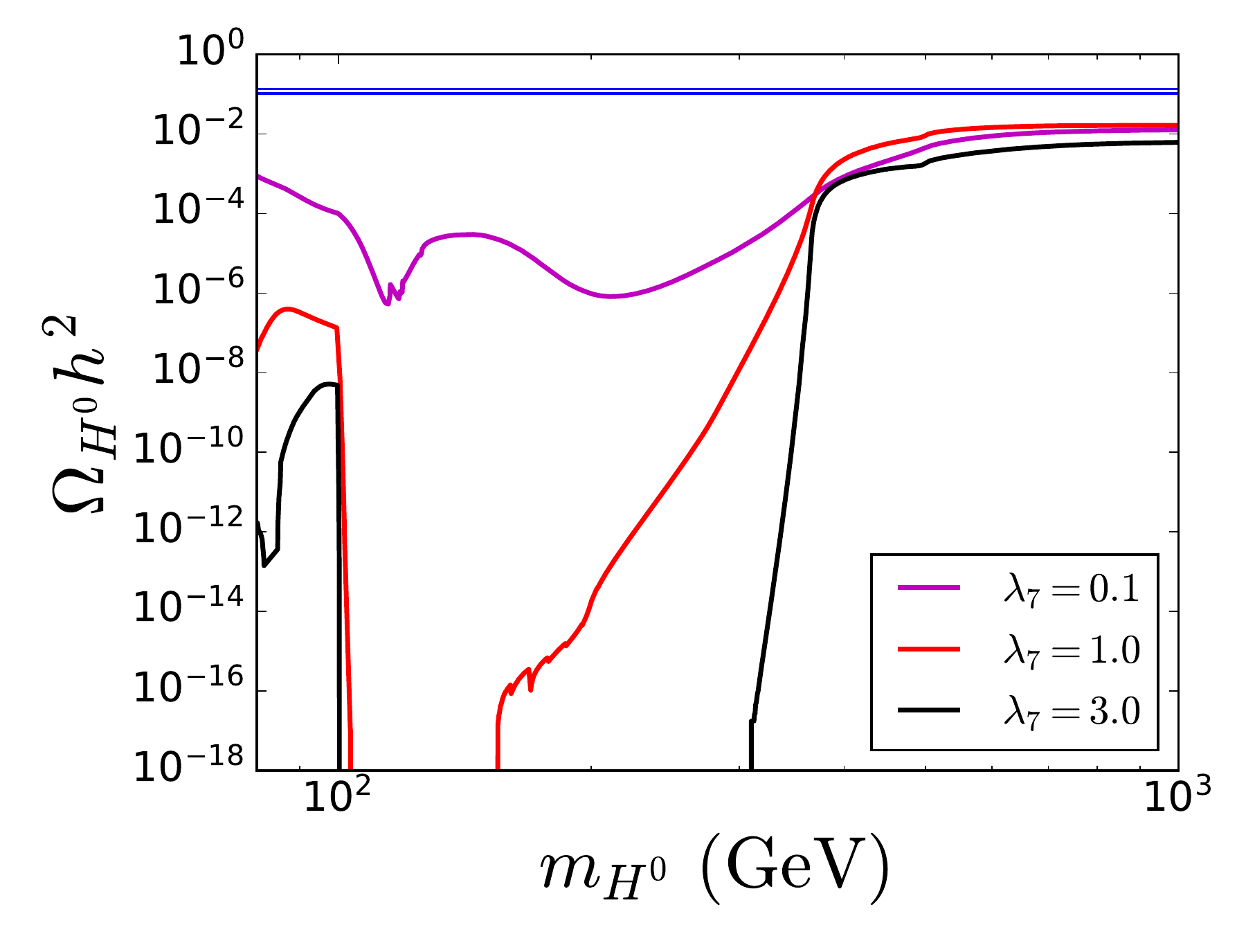}
\caption{Effect of $\lambda_7$-mediated DM semi-annihilation on $\Omega_{\phi}$ and $\Omega_{H^0}$ for two different values of  $m_{H^0}/m_\phi$: 1.2 (top panels) and  1.6 (bottom panels). In all panels $m_{H^\pm}/m_{H^0}=1.1$, $\lambda_6=0$ and $\lambda_8=\lambda_L=0.1$ have been set.}
\label{fig:relicdensity2}
\end{figure}

We first analyze the absence of semi-annihilations ($\lambda_7=0$) for the two mass hierarchies and different values of $\lambda_6$ when the DM mass ratio is fixed at $1.2$ (see Fig.~\ref{fig:relicdensity1}). In particular, we allow $\lambda_6$ to vary as $0.0,\ 1.0,\ 3.0$. In the scenario where $\phi$ is the lightest DM component (top panels), the $\lambda_6$ interaction  significantly affects the value of $\Omega_{H^0}$ (up to two orders of magnitude for intermediate values of $m_{H^0}$), while the effect on $\Omega_\phi$ is negligible. $\Omega_\phi$ is then determined by the same Higgs-mediated interactions of the scalar singlet model, resembling the singlet scalar DM model. 
When the doublet component is the lightest one (bottom panels), $\lambda_6$ deeply affects $\Omega_\phi$, with a variation of approximately three orders of magnitude over the mass range considered for $\phi$. The effect on $\Omega_{H^0}$ is, on the contrary, slightly small. In this case, $\Omega_{H^0}$ is basically determined by the Higgs and gauge portal interactions as occurs in the IDM.

\begin{figure}[t]
\centering
\includegraphics[scale=0.45]{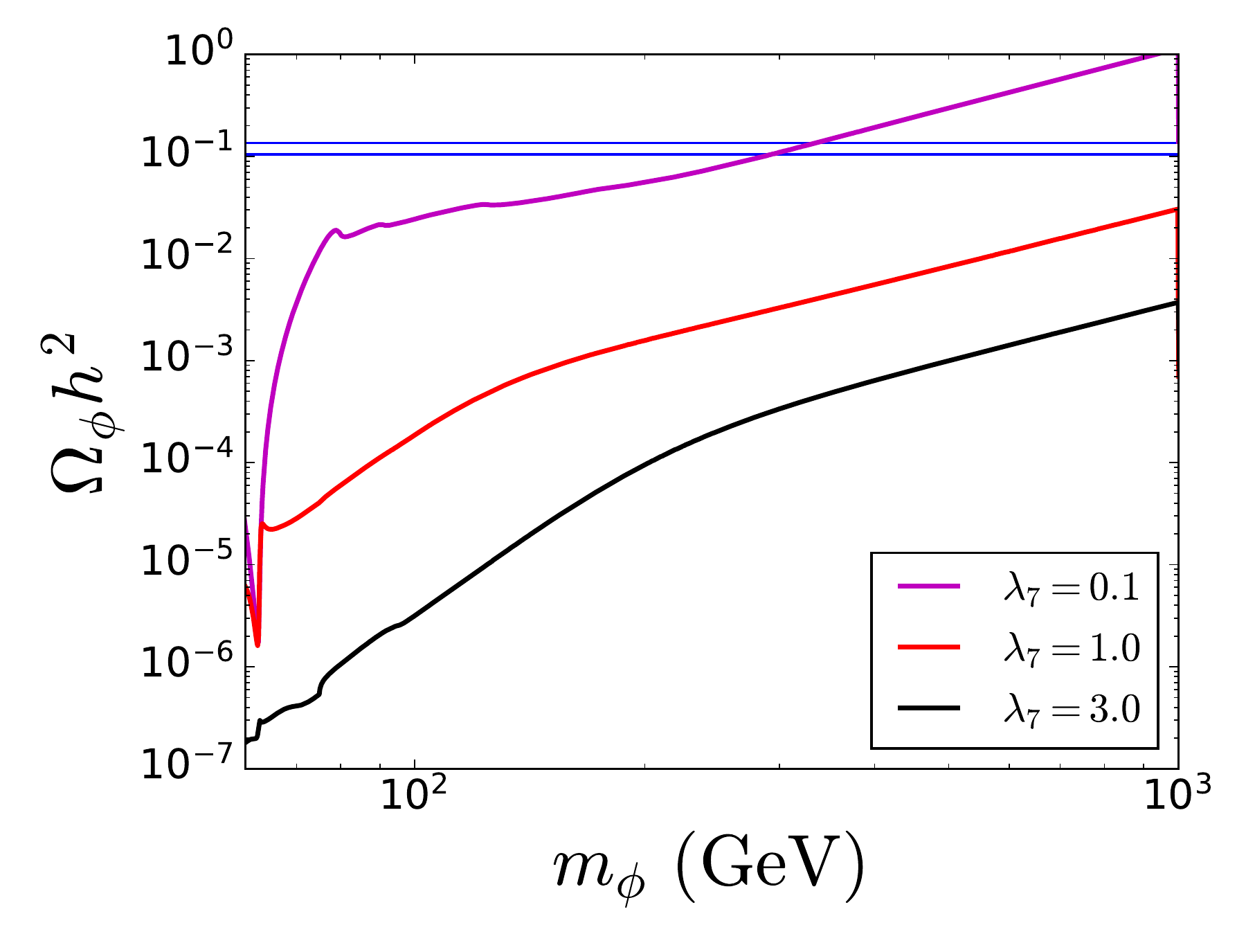}
\includegraphics[scale=0.45]{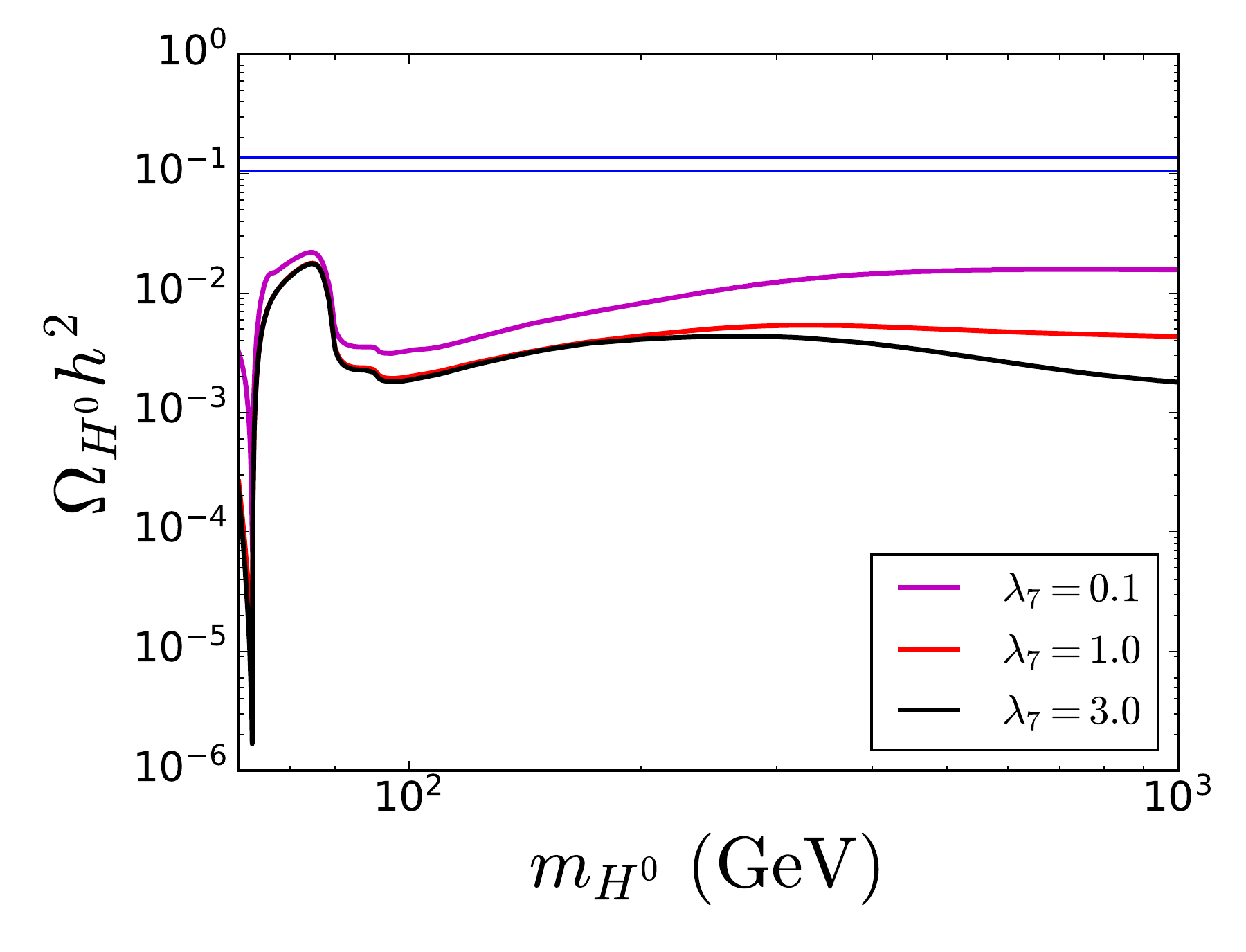}\\
\caption{Effect of $\lambda_7$-mediated DM semi-annihilations on $\Omega_{\phi}$ and $\Omega_{H^0}$ for $m_\phi/m_{H^0}=1.2$. Larger values of $m_\phi/m_{H^0}$ do not lead to significant variations of $\Omega_{\phi}$ or $\Omega_{H^0}$. We take $m_{H^\pm}/m_{H^0}=1.1$, $\lambda_6=0$ and $\lambda_8=\lambda_L=0.1$.}
\label{fig:relicdensity3}
\end{figure}

Next, we take $\lambda_6=0$ and consider the effect of the DM semi-annihilation processes when $\lambda_7$ is varied as $0.1,\ 1.0,\ 3.0$. From Fig.~\ref{fig:relicdensity2}, it can be observed that when $\phi$ is the lightest DM component, $\Omega_{H^0}$ decreases with $\lambda_7$ in several orders of magnitude over a mass range which widens with the value of mass ratio $m_{H^0}/m_{\phi}$. In the top panels, this ratio is fixed at $1.2$, whereas a value of $1.6$ is chosen for the bottom panels. This behavior is a consequence of the exponential suppression 
\begin{align}
 \frac{dn_{H^0}}{dt}\propto \sigma_v^{1210} n_\phi n_{H^0},   
\end{align}
present in the Boltzmann equation (Eq.~(\ref{eq:boltzmannEqs})) and associated with the semi-annihilation processes $\phi + H^{0*}\leftrightarrow\phi + h$ and $\phi + H^{0*}\leftrightarrow\phi + Z$ (see Fig.~\ref{fig:semi}), which do not modify the singlet abundance. This is because in the low mass region the semi-annihilation processes that alter the number density of $\phi$ are kinetically suppressed, so that $\Omega_\phi$ is governed mainly by the usual Higgs interactions. When the $\phi + \phi \rightarrow H^{0*} + h(Z)$ channels are open, $\Omega_{H^0}$ grows rapidly at the expense of a decrease of $\Omega_\phi$ of up to three orders of magnitude at the intermediate mass scale. 
Now, if $H^0$ turns out to be the lightest DM component, the impact of the $\lambda_7$ interaction on $\Omega_{H^0}$ is small, that is, instead of a large reduction like that shown in Fig.~\ref{fig:relicdensity2}, only a variation of at most one order of magnitude takes place for the different values of $\lambda_7$ on the range of $m_{H^0}$ (see Fig.~\ref{fig:relicdensity3}). Nevertheless, the DM semi-annihilation processes induced by $\lambda_7$ do significantly affect $\Omega_\phi$, although their impact is largely independent of $m_\phi/m_{H^0}$.  

As will be discussed in the next section, a suppression of at least 6 orders of magnitude on $\Omega_{H^0}$ is necessary for evading the exclusion limits on the $H^0$ SI cross section. In accordance with the discussion so far, this is possible only for a mass regime in which $m_{H^0}>m_\phi$. Therefore, hereafter we focus our analysis on the scenario where $\phi$ is the lightest DM component\footnote{The scenario $m_{H^0}>2m_\phi$ is equivalent to the complex singlet scalar DM model~\cite{Silveira:1985rk,McDonald:1993ex,Burgess:2000yq}, as the presence of $H_2$ does not affect the $\phi$ abundance because the semi-annihilation processes become inefficient.}.  
As a comment aside, we stress that the exponential suppression on $\Omega_{H^0}$ in the low mass region is effective only for the singlet-doublet two component DM in the mass regime $m_\phi<m_{H^0}$

\subsection{DM direct detection}\label{DD}  
\begin{figure}[t]
\centering
\includegraphics[scale=0.8]{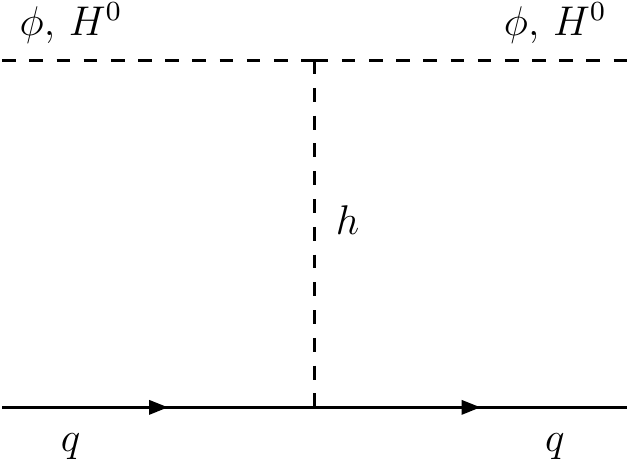}\hspace{1cm}
\includegraphics[scale=0.8]{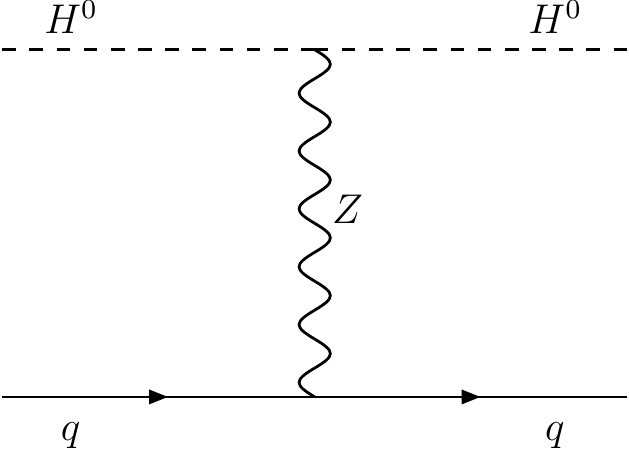}
\caption{Feynman diagrams for the elastic scattering of DM particles with nuclei mediated by the Higgs (left panel) and $Z$ bosons (right panel).}
\label{fig:DD-diags}
\end{figure}
Elastic scattering of the DM particles off nuclei can take place since both $\phi$ and $H^0$ couple to quarks by the exchange of a Higgs boson, and also by the exchange of a $Z$ boson for the case of $H^0$, as displayed in Fig.~\ref{fig:DD-diags}. 
The SI DM-nucleon scattering cross section $\sigma^\text{SI}_{\chi N}$ (for a target nucleus with atomic and mass numbers $\mathcal{Z}$ and $\mathcal{A}$) is given by~\cite{Hisano:2018bpz}
\begin{align}
    \sigma^\text{SI}_{\chi N}=\frac{1}{\pi} \left[\frac{\mu_{\chi p} \mathcal{Z} (f^S_{\chi p}+f^V_{\chi p})+\mu_{\chi n}(\mathcal{A}-\mathcal{Z})(f^S_{\chi n}+f^V_{\chi n})}{\mathcal{A}}\right]^2,
\end{align}
 where the $f^{S(V)}_{\chi \alpha}$ coefficients correspond to the scalar (vector) DM couplings to protons ($\alpha = p$) and neutrons ($\alpha = n$), whereas $\mu_{\chi\alpha}$ denotes the respective DM-nucleon reduced masses, and $\chi=\phi,H^0$. The Higgs effective couplings are given by
\begin{align}
f^S_{\chi \alpha}&=-\lambda_{\chi h}\frac{m_{\alpha} f_{\alpha}}{m_h^2 m_{\chi}},
\end{align}
with $f_{p,n}\approx 0.3$ denoting the quark content of nucleons, $\lambda_{\phi h} = \lambda_8$ and $\lambda_{H^0 h} = 2\lambda_L$. Since for $\phi$ the $Z$-mediated interaction with nucleons is not allowed, it turns out that $f_{\phi \alpha}^V=0$ for $\alpha=p,n$, whereas for $H^0$ we have
\begin{align}
f^V_{H^0\alpha}=\begin{cases}
-(1-4s_W^2)\dfrac{G_F}{\sqrt{2}},\ \alpha=p,\\
\dfrac{G_F}{\sqrt{2}},\ \alpha=n.
\end{cases}
\end{align}

Because the weak-isospin charge of $H^0$, it presents large scattering rates for typical values of the model parameters. For instance, with $|\lambda_L|<3$ and $m_{H^0}\gtrsim 100$ GeV, it is found that~\cite{Barbieri:2006dq}
\begin{align}
\sigma^{\rm{SI}}_{H^0N}&\approx\frac{G_F^2}{2\pi}\frac{\mu_N^2}{\mathcal{A}^2}\left[(\mathcal{A}-\mathcal{Z})-\mathcal{Z}(1-4s_W^2)\right]^2\approx2\times10^{-3}\,\rm{pb}.
\end{align}
To understand this, let us recall that the $Z$ couplings are proportional to $T_3\cos^2\theta_W-(Y/2)\sin^2 \theta_W$, being $\theta_W$ the Weinberg mixing angle, $T_3$ the third weak-isospin component and $Y$ the SM hypercharge. For $H^0$, $T_3=-1/2$ and $Y=1$. The fact that $Y\neq 0$ induces unsuppressed vector interactions with the $Z$ boson which give rise to an elastic SI cross-section between $H^0$ and the nucleons that is several orders of magnitude above the current upper bounds\footnote{One way to relax these constraints is to allow a mass splitting between the CP even and odd components of at least $\sim$100 keV, since this kinematically disfavors the interaction through the $Z$ portal, reducing to non-significant inelastic collisions. In the current model, however, such a scenario is not possible due to the $Z_6$ symmetry.}. 


In multi-component DM scenarios, the quantity to be compared against the DD limits provided by experimental collaborations is not merely the cross-section itself. Instead, it is the expression $\Omega_\chi \sigma^{{\rm SI}}_{\chi N}/\Omega_{{\rm DM}}$, where $\Omega_{\rm DM}$ represents the total observed DM abundance. To remain below the current upper bounds, the relic density of $H^0$ must be suppressed by at least six orders of magnitude. This suppression is precisely noticed in the mass hierarchy $m_\phi < m_{H^0}$, particularly when semi-annihilation processes are significant (see the previous section).

In order to determine the DD constraints on this two-component DM model, we determine the expected total number of events $\mathcal{N}_{\text{events}}$ in a experiment such as XENON \cite{XENON:2023cxc}, PANDAX \cite{PandaX-4T:2021bab} and LZ~\cite{LZ:2022lsv}. This can be calculated through the expression~\cite{XENON100:2011nxd}
\begin{align}
    \mathcal{N}_{\text{events}}=\omega_{exp} \int_{\mathcal{S}_{\text{min}}}^{\mathcal{S}_{\text{max}}}d\mathcal{S}\sum_{n=1}^\infty \text{Gauss}(\mathcal{S}|n,\sqrt{n}\sigma_{PMT})\int_0^\infty dE_R\epsilon(E_R)\text{Poiss}(n|\nu(E_R))\frac{dR}{dE_R},\label{Nevents}
\end{align}
where $\omega_{exp}$ is the exposure, $\mathcal{S}$ denotes the number of photo-electrons (PE) resulting from the collision between a WIMP particle and a target nucleus. $\sigma_{PMT}$ is the average single-PE resolution of the photo-multipliers arranged inside the detector to measure the scintillation photons (signal S1) resulting from the collisions. $\epsilon(E_R)$ is the detection efficiency and $\nu(E_R)$ is the expected number of PEs for a given recoil energy $E_R$. As for $dR/dE_R$, it represents the differential recoil rate per unit of detector mass. In the current scenario, this rate comes given by the sum of the rates associated with each component particle \cite{Herrero-Garcia:2017vrl, Herrero-Garcia:2018qnz}, 
\begin{align}
    \frac{dR}{dE_R}= \frac{dR_\phi}{dE_R} + \frac{dR_{H^0}}{dE_R},
\end{align}
with
\begin{align}\label{eq:differentialRate}
    \frac{dR_\chi}{dE_R}=\frac{1}{2}\frac{\rho_{\chi}\sigma_{\chi T}}{m_\chi \mu^2_{\chi T}}F^2(E_R)\mathcal{H}(E_R,m_\chi). 
\end{align}
Here $\rho_\chi\equiv (\Omega_\chi/\Omega_{{\rm DM}})\,\rho_{\odot}$ stands for the contribution of the $\chi$-component to the local DM density $\rho_{\odot}$ ($\approx 0.3$ GeV/cm$^{3}$), $\sigma_{\chi T}$ and $\mu_{\chi T}$ are, respectively, the scattering cross section and the reduced mass of the $\chi$-nucleus system, while $F^2(E_R)$ is the recoil-energy dependent nuclear form factor given by \cite{PhysRev.104.1466, LEWIN199687}
\begin{align}
    F^2(E_R)=\left[3\frac{j_1(qR)}{qR}\right]^2e^{-q^2s^2},
\end{align}
where $j_1$ is the spherical Bessel function of the first kind, $q=\sqrt{2m_T E_R}$\ \ and $R = \sqrt{c^2+\frac{7}{3}\pi^2a^2-5s^2}$, with $c = (1.23 A^{1/3}-0.6)$ fm, $a=0.52$ fm and $s = 0.9$ fm. On the other hand,
\begin{align}
    \mathcal{H}(E_R,m_\chi) = \int_{v_\text{min}}^\infty \frac{f_{\oplus}(\mathbf{v}_\text{rel})}{v_\text{rel}}d^3\mathbf{v}_\text{rel},
\end{align}
$f_{\oplus}$ being the astrophysical DM velocity distribution measured with respect to the lab frame and $v_\text{min}$ the minimum speed needed to produce a recoil with energy $E_R$
\begin{align}
    v_\text{min}(E_R,m_\chi)=\sqrt{\frac{(m_{\chi}+m_T)^2E_R}{2m^2_{\chi}m_T}}.
\end{align}
 Regarding to the galactic frame, and assuming the so-called Standard Halo Model~\cite{PhysRevD.33.3495, PhysRevD.37.3388}, these velocities follow a Maxwell-Boltzmann distribution of the form
\begin{align}
    f(\mathbf{v})=\begin{cases}\frac{1}{N}e^{-|\mathbf{v}|^2/v^2_0},\ \text{for}\ \ |\mathbf{v}|<v_{\mathrm{esc}},\\
    0,\ \ \text{for}\ \ |\mathbf{v}|>v_{\mathrm{esc}}, \end{cases}
\end{align}
with galactic escape velocity $v_{\mathrm{esc}}$ ($\sim 540$ km/s), velocity dispersion $v_0$ ($\sim 220$ km/s) \cite{XENON:2018voc}\footnote{For simplicity, we assume that these parameters are the same for both DM components. A more general analysis with different dispersion velocities was carried out in Refs.~\cite{Herrero-Garcia:2017vrl, Herrero-Garcia:2018qnz}.} and normalization 
\begin{align}
    N=\pi^{3/2}v^3_0\left[ \mathrm{erf}\left(\dfrac{v_{\mathrm{esc}}}{v_0}\right)-\dfrac{2v_{\mathrm{esc}}}{\sqrt{\pi}v_0}e^{-\left(\frac{v_{\mathrm{esc}}}{v_0}\right)^2} \right].
\end{align}
Thus, if $\textbf{v}_E$ is the Earth's velocity with respect to the galactic frame ($\sim 232$ km/s \cite{XENON:2018voc}), then $f_{\oplus}(\mathbf{v}_\text{rel})=f(\mathbf{v}_\text{rel}+\mathbf{v}_E)$.
Finally, due to the scalar nature of the DM candidates, the particle-physics dependent observable $\sigma_{\chi T}$ can be written in terms of the SI $\chi$-nucleon scattering cross section $\sigma^\text{SI}_{\chi N}$ as
\begin{align}
    \sigma_{\chi T}=\left(\mathcal{A}\frac{\mu_{\chi T}}{\mu_{\chi N}}\right)^2\sigma^\text{SI}_{\chi N}.\label{eq:sigmaDMT}
\end{align}

Putting it all together, we can express the number of events predicted by the model in DD experiments as the sum of the events induced by the singlet and those generated by the doublet, i.e.,
\begin{align}
    \mathcal{N}_\text{events} = \mathcal{N}^{\phi}_\text{events} + \mathcal{N}^{H^0}_\text{events}, 
\end{align}
where each contribution is proportional to the product of the corresponding local relic density $\rho_\chi$ and the SI scattering cross section with nucleons $\sigma^\text{SI}_{\chi N}$, as indicated in Eqs. (\ref{eq:differentialRate}) and (\ref{eq:sigmaDMT}). 

\section{Numerical analysis}\label{sec:Numerics}
In this section we investigate the phenomenological implications of the constraints on the model and establish the regions of the parameter space where semi-annihilation processes render it viable. 
\subsection{Theoretical constraints}\label{sec:Constraints}
\subsubsection{Perturbativity}\label{sec:perturbativity}
To guarantee that tree-level corrections are always more relevant than the one-loop contributions, we demand vertex factors to be less than $4\pi$ in the Feynman rules associated with the quartic interactions \cite{Lerner:2009xg}.  If several upper bounds for a same coupling are possible, the most stringent one is chosen. The corresponding upper bounds are given by
\begin{align}
\lambda_{2}<\frac{2\pi}{3},\ |\lambda_3|<4\pi,\ |\lambda_3+\lambda_4|<4\pi,\ |\lambda_4|<8\pi,\ |\lambda_8|<4\pi,\ |\lambda_6|<4\pi,\ |\lambda_{\phi}|<\pi.
\end{align}
\subsubsection{Perturbative unitarity}\label{sec:unitarity}
When the energy involved in a scalar-scalar scattering process is high enough in comparison with the masses of the involved particles, all the contributions to the tree-level scattering matrix mediated by propagators are negligible so that only quartic point interactions are relevant. In this limit the $s$-wave scattering amplitudes must satisfy the perturbative unitarity condition, which results in the fact that the eigenvalues of the scattering matrices must all be less than 8$\pi$~\cite{Goodsell:2018tti}. The unitarity bounds for the current model can be obtained from the general analysis reported in Ref.~\cite{Belanger:2014bga}.  When the initial scattering states are classified according to the total hypercharge $Y$, weak isospin $T_3$ and discrete $Z_6$ charge, the $S_{T_3}^{Y}$ scattering matrices can be expressed as
\begin{gather}
    8\pi S^{2}_{1}=\begin{pmatrix}
        2\lambda_1 &0 &0\\
        0 &2\lambda_2 &0\\
        0 &0 &\lambda_3 + \lambda_4
    \end{pmatrix},\ \ 
    8\pi S^{2}_{0}=\lambda_3-\lambda_4,\\ 
    8\pi S^{1}_{1/2}=\begin{pmatrix}
        \lambda_6 &\lambda_7 &0 &0\\
        \lambda_7 &\lambda_8 &0 &0\\
        0        &0 &\lambda_8 &0\\
        0        &0 &0         &\lambda_6
    \end{pmatrix},\ \ 
    8\pi S^{0}_{1}=\begin{pmatrix}
        2\lambda_1 &\lambda_4 &0 &0\\
        \lambda_4  &2\lambda_2 &0 &0\\
        0          &0 &\lambda_3 &0\\
        0          &0 &0         &\lambda_3  
    \end{pmatrix},\ \
    8\pi S^{0}_{0}=\begin{pmatrix}
         A_{3\times 3} &0_{3\times 4}\\
         0_{4\times 3}  &B_{4\times 4}
    \end{pmatrix},
\end{gather}
where
\begin{equation}
  A_{3\times 3}=\begin{pmatrix}
        6\lambda_1  &2\lambda_3+\lambda_4  &\sqrt{2}\lambda_8 \\
        2\lambda_3+\lambda_4  &6\lambda_2  &\sqrt{2}\lambda_6 \\
        \sqrt{2}\lambda_8  &\sqrt{2}\lambda_6  &\lambda_\phi  
    \end{pmatrix},\ \
  B_{4\times 4}=\begin{pmatrix}
        \lambda_\phi  &0  &\lambda_7  &0\\
        0  &\lambda_\phi  &0  &\lambda_7\\
        \lambda_7  &0  &\lambda_3+2\lambda_4  &0\\
        0  &\lambda_7  &0  &\lambda_3+2\lambda_4
    \end{pmatrix}.
\end{equation}  

\subsubsection{Vacuum stability}\label{sec:stability}
Vacuum stability demands a scalar potential bounded from below. To establish the corresponding conditions, we consider as usual high field values so that only quartic terms are relevant, and build the quartic interaction matrix in terms of non-negative field variables in the following way \cite{Ginzburg:2004vp}: 
\begin{align}
    |H_1|^2=r_1^2,\ |H_2|^2=r_2^2,\ H_1^\dagger H_2 = r_1r_2\rho e^{i\theta},\ \phi = r_\phi e^{i\theta_\phi}.   
\end{align}
Here $r_1=r\cos\gamma$ and $r_2=r\sin\gamma$, with $r,r_\phi\geq 0$ and $0\leq\gamma\leq \pi/2$, whereas $|\rho|\leq 1$ and $\theta,\theta_\phi \in [0, 2\pi]$. The necessary and sufficient stability conditions are obtained by demanding this matrix to be copositive \cite{Kannike:2012pe}. For the $Z_6$ model discussed here, the conditions are \cite{Belanger:2014bga}
\begin{align}
\lambda_{\phi,1,2}>0,\ \lambda_3+2\sqrt{\lambda_1\lambda_2}>0,\ \lambda_3+\lambda_4+2\sqrt{\lambda_1\lambda_2}>0,\ \sqrt{\Lambda_{11}\Lambda_{22}}+\Lambda_{12}>0, 
\end{align}
with
\begin{align}
\Lambda_{11}&=\lambda_{1}\cos^4\gamma + \left(\lambda_3+\lambda_4\rho^2\right)\cos^2\gamma\sin^2\gamma+\lambda_{2}\sin^4\gamma,\\
\Lambda_{22}&=\lambda_\phi,\\
\Lambda_{12}&=\frac{1}{2}\left[\lambda_8\cos^2\gamma+\lambda_7\rho\cos\gamma\sin\gamma\cos(\theta-2\theta_\phi) + \lambda_6\sin^2\gamma\right].
\end{align}
These conditions must be fulfilled for all defined values of $\rho,\ \gamma,\ \theta$ and $\theta_\phi$.  

\subsubsection{Renormalization group equations}
We determine the evolution with energy of all dimensionful and dimensionless parameters by solving the two-loop renormalization group equations (REGs) obtained with {\tt SARAH}~\cite{Staub:2013tta, Staub:2015kfa}, and checking that the conditions of perturbative unitarity and vacuum stability, as well as perturbativity are satisfied for the different values of the energy scale $\Lambda$. 
To determine the viable parameter space we only consider points for which all the theoretical conditions are guaranteed up to energy scales above the highest mass present in the dark sector, which is $m_{H^\pm}$ in the present setup.

\subsection{Experimental constraints}
\subsubsection{EW precision tests}\label{sec:stu}
The new scalar fields may modify the vacuum polarization of the gauge bosons. These effects are parameterized by the so-called EW oblique parameters $S$, $T$ and $U$~\cite{Peskin:1991sw}. The SM best fit reads $\bar S=0.06\pm 0.09$, $\bar T=0.10\pm 0.07$ with a correlation coefficient $+0.91$ (under the assumption $U=0$)~\cite{Baak:2014ora}. 
In the present model the new contributions to $S$ and $T$ are given by \cite{Barbieri:2006dq}
\begin{align}
S = -\frac{\ln\left(r\right)}{6\pi},\quad T = \frac{m_{H^0}^2}{16\pi m_W^2 s_W^2}\frac{r^4-1-4r^2\ln r}{r^2-1},
\end{align}
where $r=m_{H^\pm}/m_{H^0}$, $m_W$ is the $W$ boson mass and $s_W^2=0.223$. 
An agreement with the EW precision tests is maintained as long as the splitting between the masses of the $H^0$ and $H^\pm$ is small.

\subsubsection{Collider constraints}
For the diphoton channel, the signal strength $R_{\gamma \gamma}$ measures the ratio of the observed diphoton production cross section relative to the SM expectation~\cite{Posch:2010hx}: 
\begin{align}
R_{\gamma \gamma} &= \dfrac{\sigma(pp \to h \to \gamma \gamma)^{\rm{Z_6}}}{\sigma(pp \to h \to \gamma \gamma)^{\rm{SM}}} 
            = \dfrac{\sigma(pp \to h \to \gamma \gamma)^{\rm{IDM}}}{\sigma(pp \to h \to \gamma \gamma)^{\rm{SM}}} 
                          \approx \dfrac{[{\rm Br}( h  \to \gamma \gamma) \big ] ^{\rm{IDM}}}{[{\rm Br}( h \to \gamma \gamma)\big ]^{\rm{SM}}}~.
\end{align}
This observable was measured by ATLAS~\cite{ATLAS:2019jst} and CMS~\cite{CMS:2021kom} with 139 fb$^{-1}$ obtaining $R_{\gamma \gamma}^{\rm ATLAS}  = 1.03\pm 0.12,$ and $ R_{\gamma \gamma}^{\rm CMS}  = 1.12 \pm 0.09$. 
The impact of the $Z_6$-odd charged scalars over the decay ratio can be quantified from Ref.~\cite{Abe:2014gua} as
\begin{align}
&R_{\gamma\gamma}= \left|1 + \frac{1}{A_{SM}}\left[\frac{\lambda_3 v^2\, A_S(\tau_{H^\pm})}{2 m^2_{H^\pm}}\right]  \right|^2,
\end{align}
where $A_{{\rm SM}}=-6.5$ is the SM contribution from charged fermions and gauge bosons, $\tau_{H^\pm}=m^2_{h}/(4 m^2_{H^\pm})$ and $A_S(\tau)=-[\tau-\arcsin^2(\sqrt{\tau})]/\tau^{2}$.  %

The LHC also sets bounds on the invisible Higgs decays. If one or both DM particles are lighter than half the Higgs mass, the $h\to \chi^*\chi$ decay would be allowed, contributing to the invisible branching ratio of the Higgs boson, $\mathcal{B}_{inv}$. The decay width associated with these processes is given by
\begin{align}
\Gamma(h\to \chi^*\chi)&=\frac{\lambda^2_{\chi}v^2}{16\pi m_h}\left[1-\frac{4m^2_{\chi}}{m_h^2}\right]^{1/2},
\end{align}
with $\lambda_\chi=\lambda_{8}$ for $\chi=\phi$, and $\lambda_\chi=2\lambda_L$ for $\chi=H^0$. To be consistent with current data, we require $\mathcal{B}_{inv}\leq 0.13$ \cite{Sirunyan:2018owy,ATLAS:2020cjb}.

On the other hand, LEP sets limits on the masses of all charged particles which can be directly produced, as well as on particles resulting from their decays. These limits can be easily reinterpreted for the new scalars present in the model. The decays of gauge bosons into $Z_6$-odd pairs are excluded by their invisible width measurements~\cite{ParticleDataGroup:2020ssz}, leading to the constraints 
\begin{align}
m_{H^0}+m_{H^\pm}> m_{W},\ \ m_{H^0,H^\pm} > \frac{m_{Z}}{2}.    
\end{align}
Direct chargino searches at LEP II can also be reinterpreted for the search of charged scalars \cite{Pierce:2007ut}, leading to $ m_{H^\pm} > 70$ GeV. For a compressed spectra, a tighter constraint applies, namely $m_{H^\pm} > 100$ GeV.

\subsubsection{Dark matter constraints}
Since both $\phi$ and $H^0$ constitute all the DM present in the universe, their abundances must satisfy the condition
\begin{align}
\Omega_\phi +\Omega_{H^0} =\Omega_{{\rm DM}},
\end{align}
with $\Omega_{{\rm DM}}$ denoting the total observed DM abundance as measured by the PLANCK collaboration \cite{Planck:2018vyg} $\Omega_{{\rm DM}}h^2=0.1198\pm 0.0012$.  
Given that the theoretical prediction on the relic density is not expected to be as precise as that of PLANCK, in our scans we consider a model compatible with that value if the DM abundance, as determined by {\tt MicrOMEGAs}, lies between 0.11 and 0.13, which amounts to about a 10\% uncertainty. In what follows, we denote the fraction of the total DM density accounted for by each component as 
\begin{align}
\xi_\chi \equiv\frac{\Omega_\chi}{\Omega_{{\rm DM}}},\ \ \chi=\phi,H^0,
\end{align}
so that $\xi_\phi +\xi_{H^0} = 1$. 

Given that currently the strongest DD constraints are those reported by the LZ collaboration with an exposure of $60.0$ days and a fiducial mass of $5.5$ ton of liquid xenon ($\mathcal{A}=131$, $\mathcal{Z}=54$ and $m_T=122.0$ GeV) \cite{LZ:2022lsv}, we consider the characteristics and specifications of this detector to determine $\mathcal{N}_\text{events}$ from Eq.(\ref{Nevents}). For this goal, we take $\mathcal{S}$ between 3 and 80 PE \cite{LZ:2022lsv} and fix $\sigma_{PMT} = 0.4$ according to Ref.~\cite{Baudis:2013xva}, whereas $\epsilon(E_R)$ is read from the black solid line of Fig. 2 in Ref.~\cite{LZ:2022lsv}. From the $S_1$ yield given in the upper left panel of Fig. 2 in Ref.~\cite{LZ:2018qzl} we can extract $\nu(E_R)$ since this corresponds to $\nu(E_R)/E_R$.  

With the aid of a Test Statistic (TS), it is possible to obtain an upper bound for $\mathcal{N}_\text{events}$. Closely following Ref.~\cite{Cirelli:2013ufw}, we take
\begin{align}
    \text{TS}(m_\chi)=-2\ln \left[\frac{\mathcal{L}(\mathcal{N}_\text{events})}{\mathcal{L}_\text{BG}}\right],
\end{align}
with
\begin{align}
    \mathcal{L}(\mathcal{N}_\text{events}) = \frac{(\mathcal{N}_\text{events}+\mathcal{N}_{BG})^{\mathcal{N}_\text{obs}}}{\mathcal{N}_\text{obs}!}e^{-(\mathcal{N}_\text{events}+\mathcal{N}_\text{BG})},
\end{align}
and $\mathcal{L}_\text{BG}\equiv \mathcal{L}(0)$. Here $\mathcal{N}_\text{obs}$ and $\mathcal{N}_\text{BG}$ stand for, respectively, the number of observed and background events. By requiring $\text{TS}(m_\chi)>2.71$, 90\% CL limits can be obtained. With $\mathcal{N}_\text{obs}=0$ and $\mathcal{N}_\text{BG}= 333$ \cite{Aprile:2018dbl}, the expected number of events must fulfill $\mathcal{N}_\text{events} \lesssim 31$. This bound will be used to constraint the free parameters of the model. 

Furthermore, to gain insight into the parameter space regions within the projected sensitivities of planned experiments, we consider LZ with a full exposure $\omega_\text{exp}=15.33$ ton$\cdot$yr \cite{Mount:2017qzi} and apply the maximum gap method assuming zero observed events \cite{PhysRevD.66.032005}. In this hypothetical scenario, we require $1-\text{exp}(-\mathcal{N}_\text{events})\geq 0.9$, which implies $\mathcal{N}_\text{events} \lesssim 2.3$.

\subsection{Scan}\label{sec:scan}
To explore the viability of the model, we conducted a random scan over the new particle masses and the free scalar couplings and determined the parameter space regions in which all constraints imposed on the model are satisfied. Specifically, these parameters are allowed to randomly vary as
\begin{align}
   &40\ \text{GeV} \leq m_\phi \leq 1000\ \text{GeV},\quad m_\phi < m_{H^0} < 2m_\phi,\quad m_{H^0} < m_{H^\pm} \leq 2\, {\rm TeV},\\
   &10^{-4}\leq \lambda_2<\frac{2}{3}\pi,\quad  10^{-4}\leq \lambda_\phi<\pi,
   \quad   10^{-4}\leq |\lambda_3|,|\lambda_8|,|\lambda_6|,|\lambda_7|<4\pi. 
\end{align}
Some of the observables, as DM abundances, rates of ID processes and the Higgs invisible decay width, were numerically determined by means of the {\tt MicrOMEGAs} whereas others were calculated analytically. 

\subsection{Results}

\begin{figure}
\centering
\includegraphics[scale=0.45]{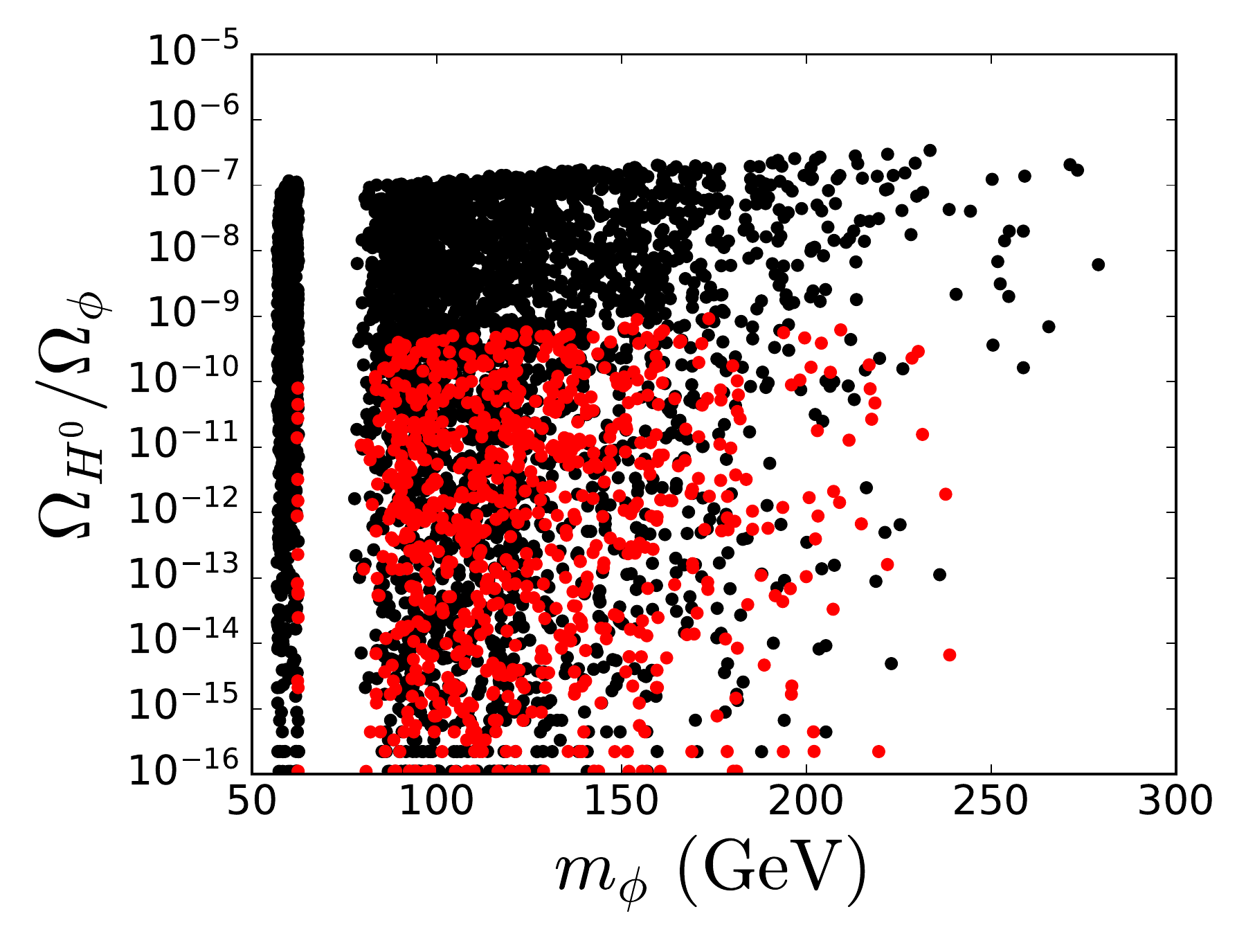}
\includegraphics[scale=0.45]{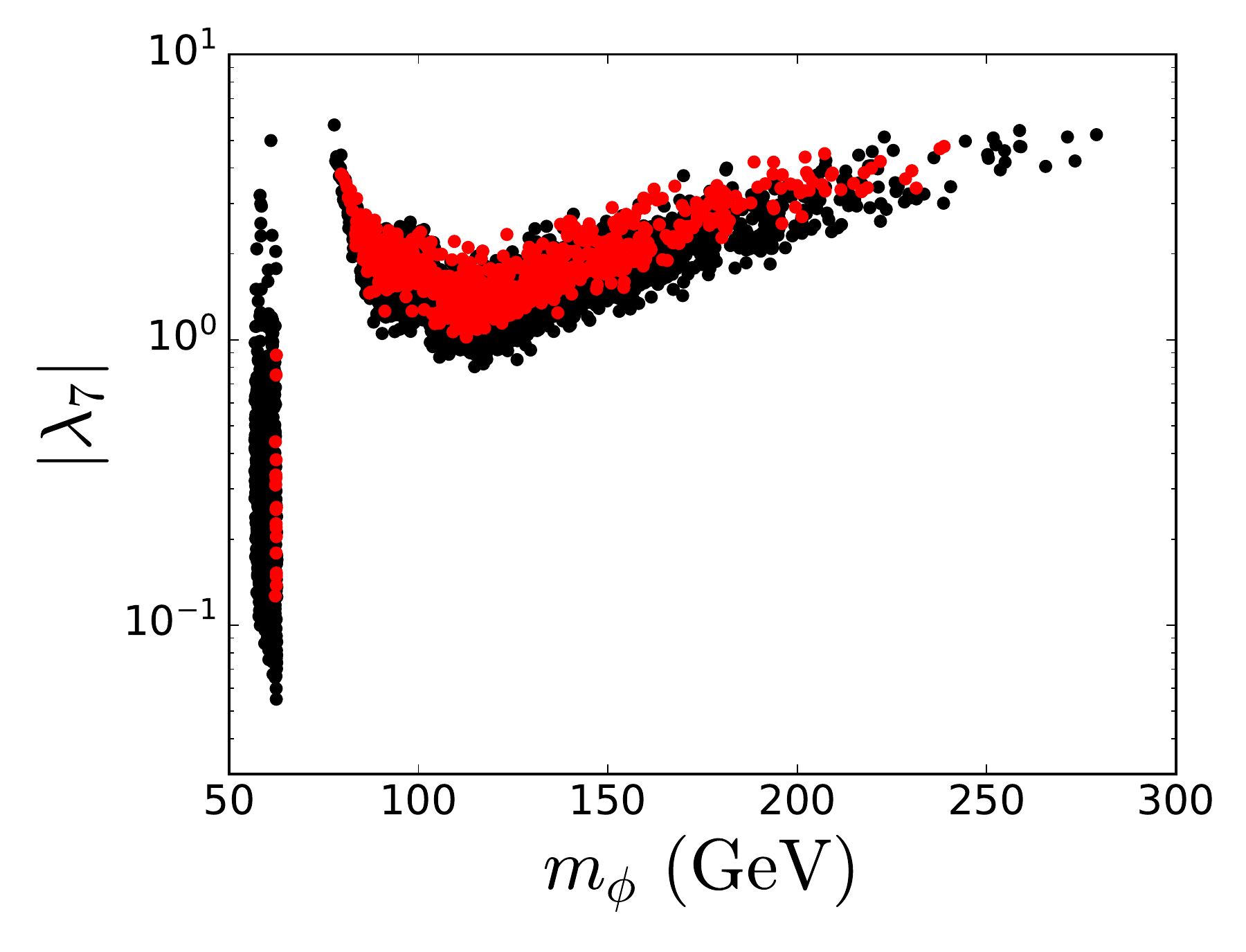}\\
\includegraphics[scale=0.45]{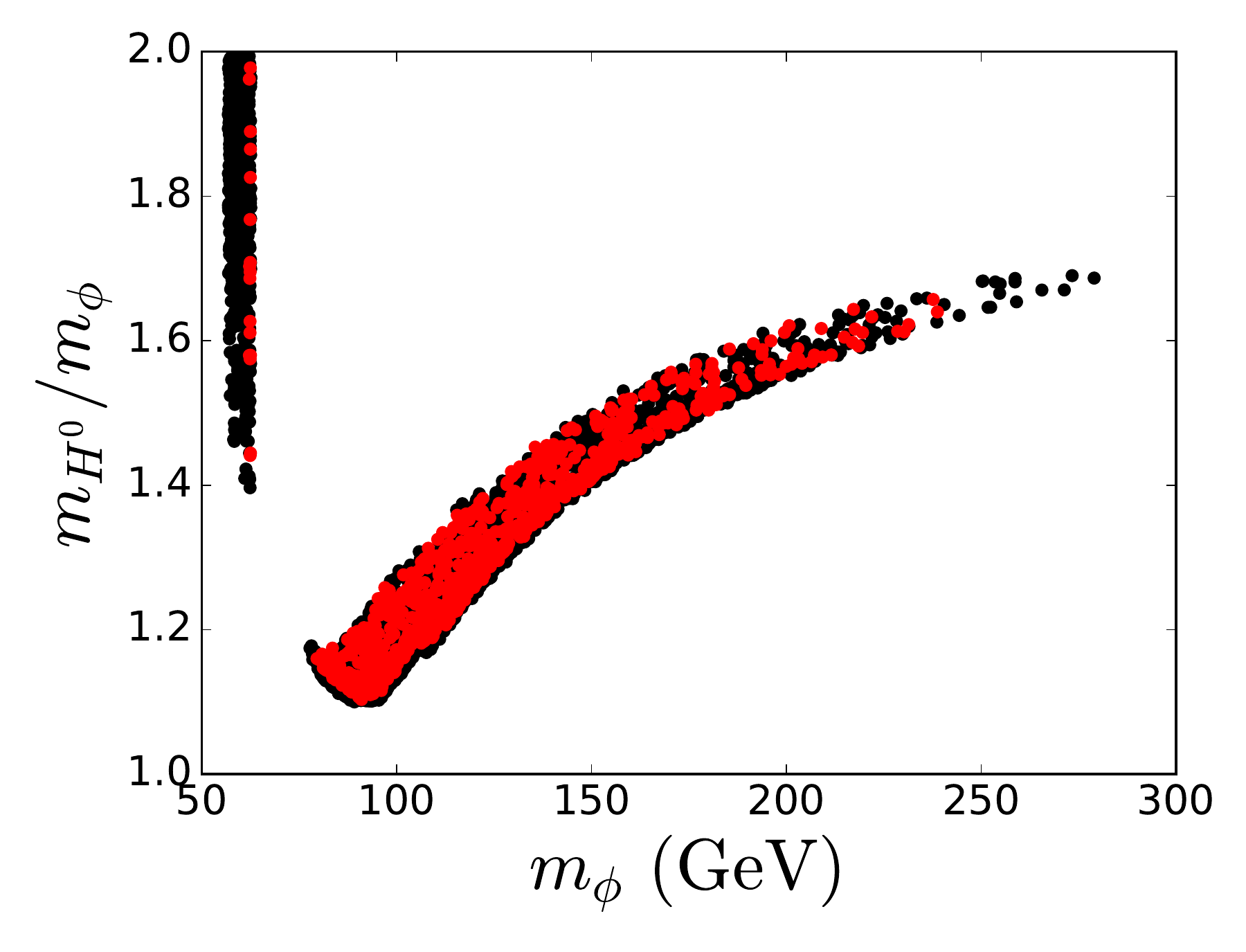}
\includegraphics[scale=0.45]{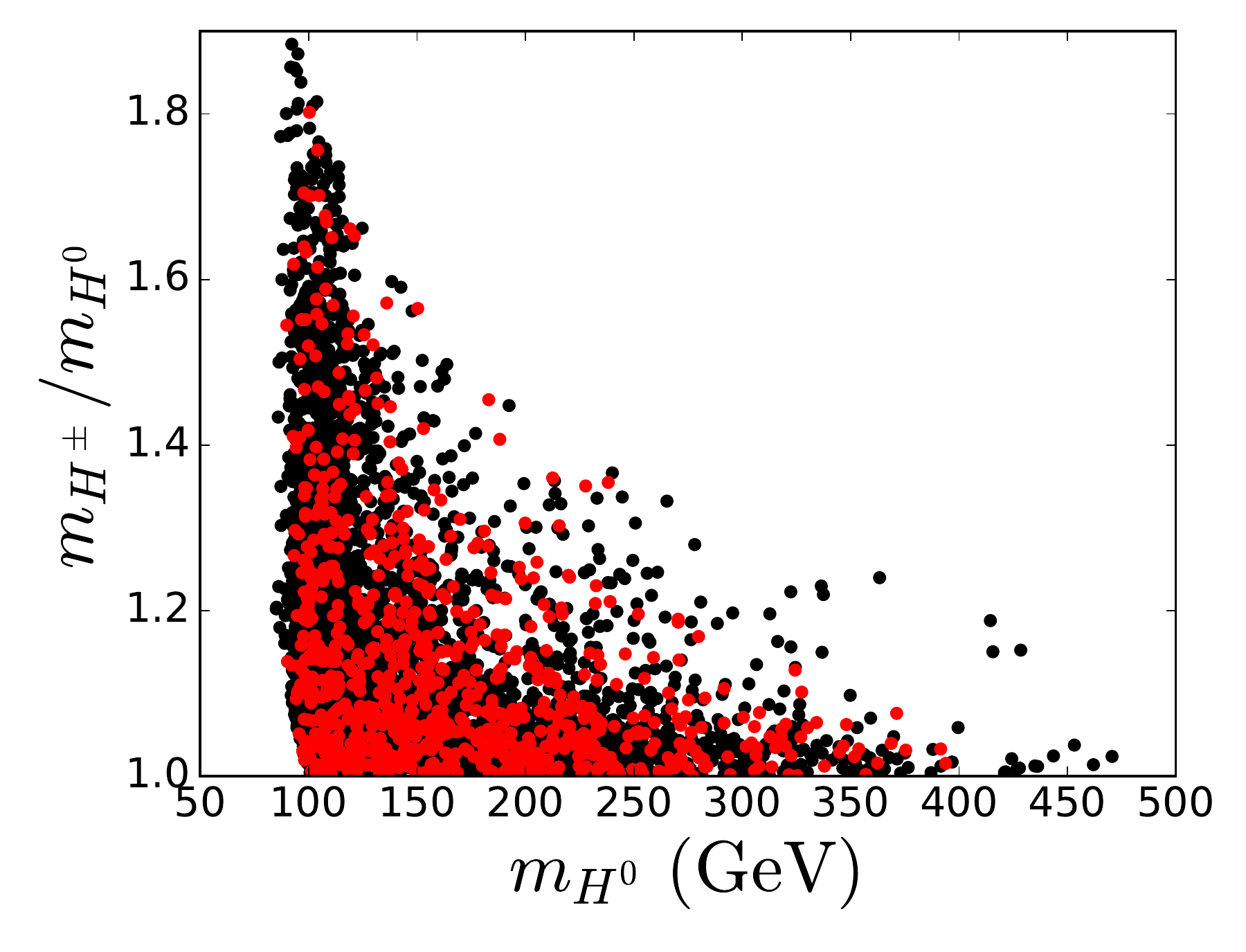}\\
\includegraphics[scale=0.45]{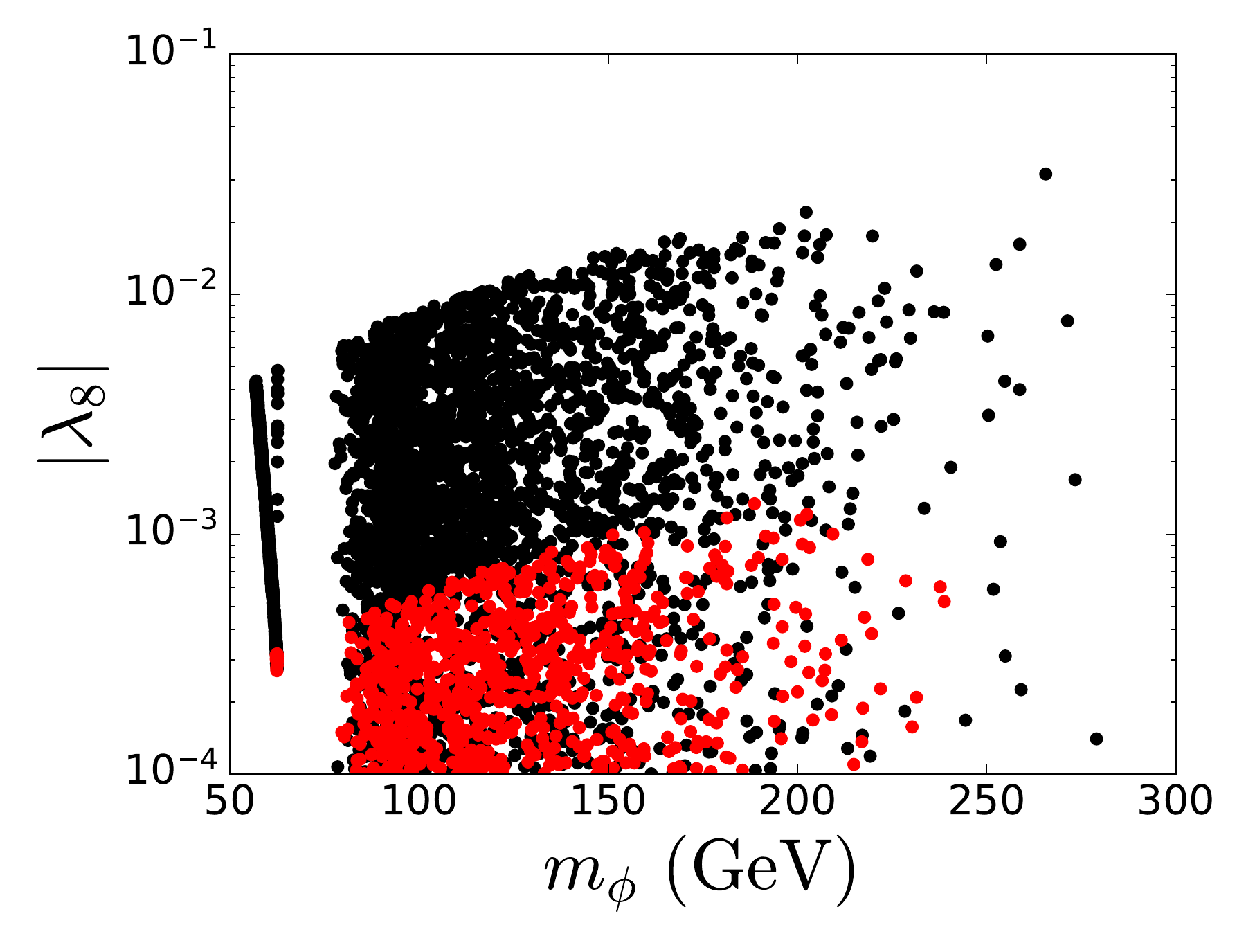}
\includegraphics[scale=0.45]{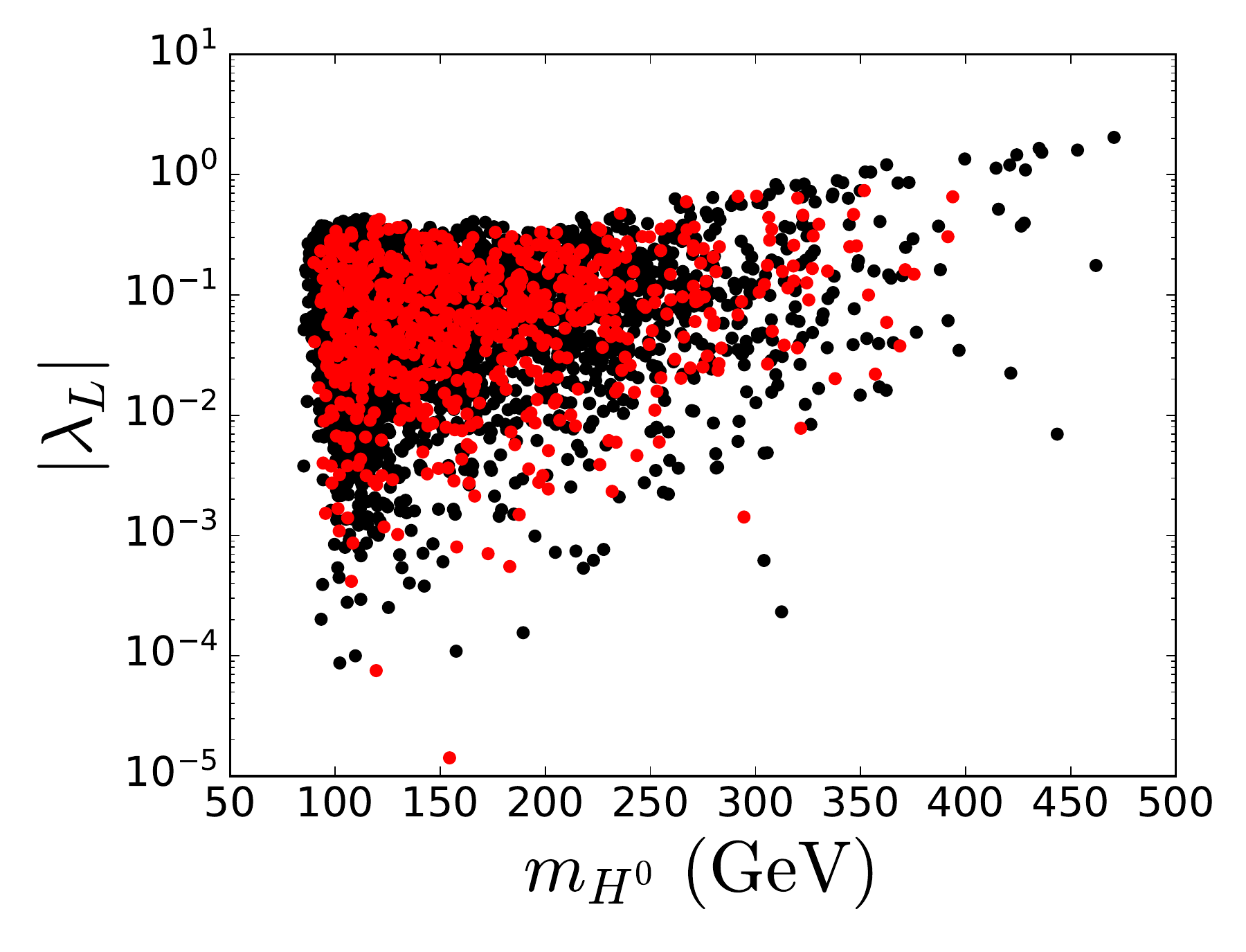}
\caption{Parameter space of the model. On the left (right), the mass ratio $m_{H^0}/m_\phi$ ($m_{H^\pm}/m_{H^0}$) is shown as a function of the $\phi$ ($H^0$) mass. The ratio of the DM relic abundances as a function of $m_\phi$ also is displayed. Scalar couplings $\lambda_7$ and $\lambda_8$ ($\lambda_L$) are shown in terms of the $\phi$ ($H^0$) mass. }
\label{fig:parspace}
\end{figure}

The resulting viable parameter space projected onto different planes is displayed in Fig.~\ref{fig:parspace}. The red points correspond to those where the upper bound on the expected number of events in the LZ experiment is satisfied, while the black points denote those that would be excluded with a full exposure of 15.33 t$\cdot$y if no events are observed.  
The most significant conclusion drawn from this figure is that, thanks to the semi-annihilation processes, it is possible to satisfy stringent DD bounds with DM masses around the EW scale.  
As can be seen from the top panels, $\lambda_7$ must be large enough ($\lambda_7\gtrsim 1$) to provide the sufficient suppression on $\Omega_{H^0}$ (at least seven orders of magnitude with respect to $\Omega_\phi\approx\Omega_{{\rm DM}}$)\footnote{$\lambda_6$ is hardly constrained meaning that the conversion processes are not relevant on setting the DM relic abundances.} and lead to a small contribution to $\Omega_\text{DM}$. In this way, for the setup considered in this work, the lightest component $\phi$ constitutes the bulk of the DM content of the universe.   

The center panels show that the viable models involve DM masses ranging from the Higgs resonance up to approximately 280 GeV for the singlet component (left panel), and from just over 90 GeV to around 480 GeV for the doublet component (right panel). The low mass region ($\sim 40$ GeV) is excluded by Higgs and collider physics. It is important to highlight that  the DM masses do not need to be degenerated.   
As for the charged and neutral doublet components, the mass splitting between them is strongly constrained by the EW precision tests and decreases with $m_{H^0}$ (right panel). 

The Higgs portal couplings with the singlet ($\lambda_8$) and the doublet ($\lambda_L$) are displayed in the bottom panels. $\lambda_L$ is not as constrained as $\lambda_8$, as it can take values as small as the lower bound defined in the scan ($10^{-4}$) and as large as $\sim 1$. This is because, although the $H^0$-nuclei scattering cross-section depends explicitly on $\lambda_L$, it is completely dominated by the $Z$ portal, rendering irrelevant the effect of the Higgs portal in such a case. 
In contrast, $\lambda_8$ is constrained to below $0.01$ due to its effect on the expected events in LZ.  This upper limit would be tightened by almost an order of magnitude under the most conservative scenario of the future LZ setup. 
It is worth noting that the expected number of events in DD experiments has two contributions, one due to $\phi$ and another to $H^0$, both proportional to the product of their respective local relic densities and scattering cross-sections with nuclei. Thus, to fulfill a more stringent future bound on $\mathcal{N}_\text{events}$, a decrease of the magnitude of $\lambda_8$ is necessary, thus reducing the event fraction associated with $\phi$.  
  
\begin{figure}
\centering
\includegraphics[scale=0.45]{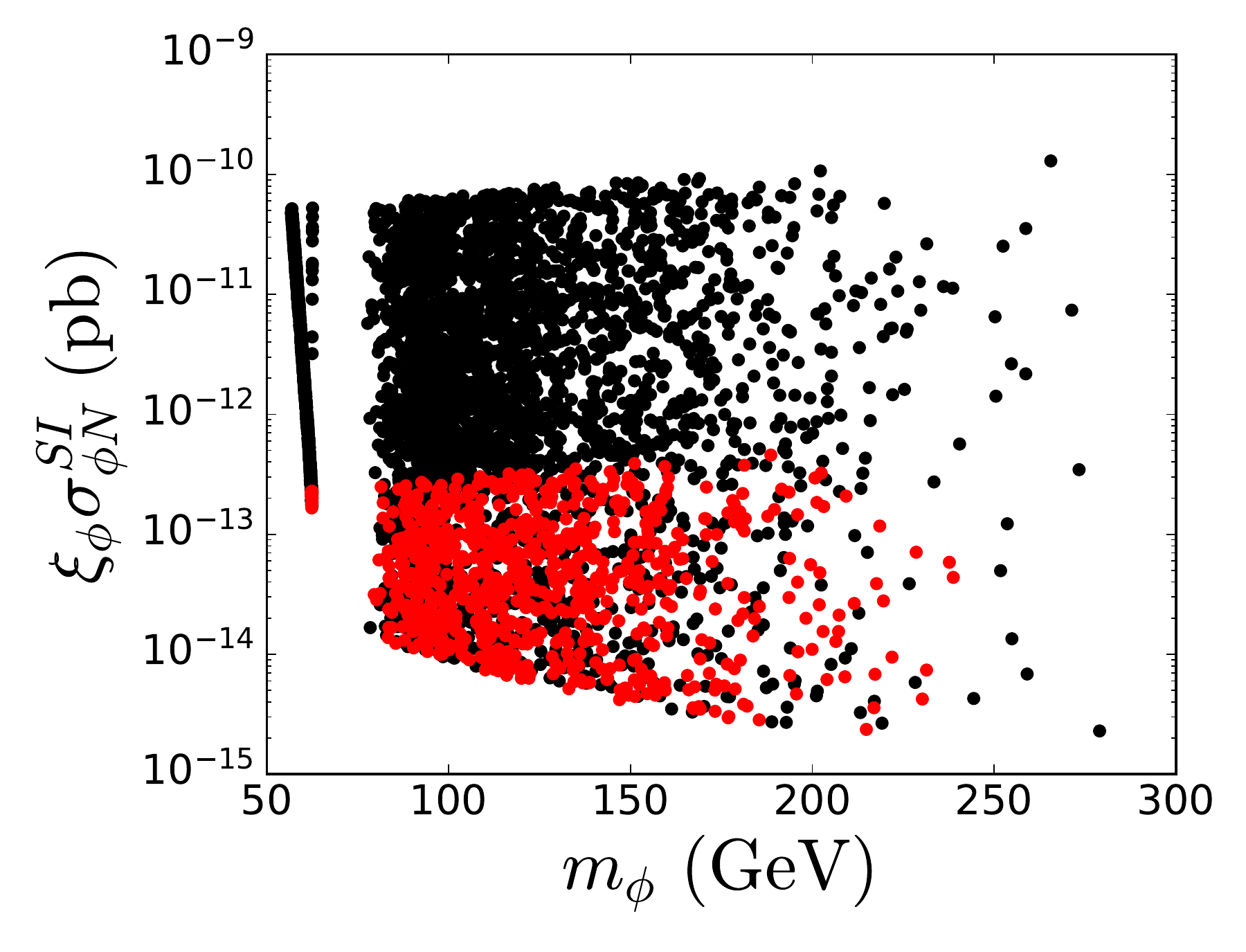}
\includegraphics[scale=0.45]{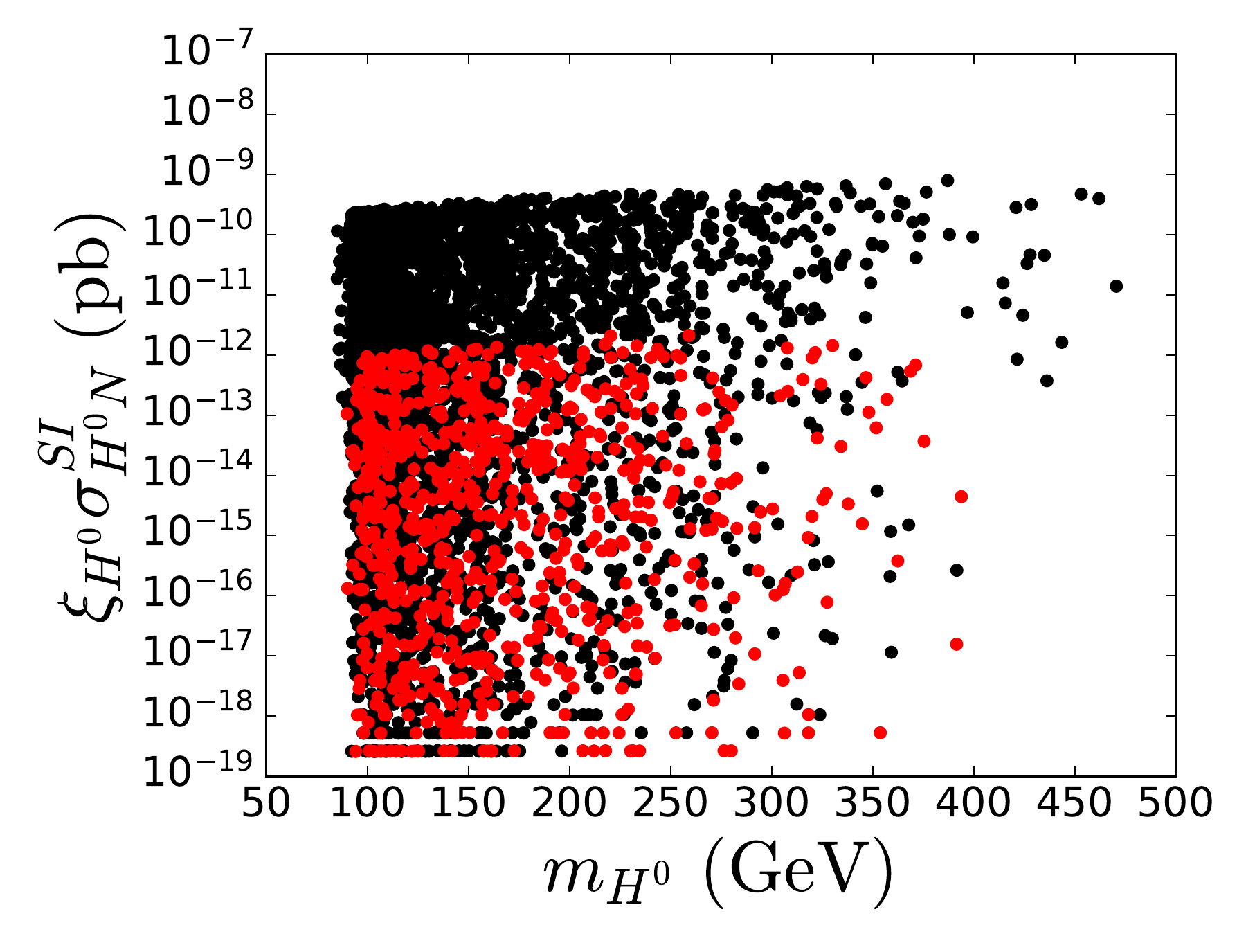}
\caption{SI cross-sections for elastic scattering of $\phi$ (left panel) and $H^0$ (right panel) with nuclei scaled by $\xi_\phi$ and $\xi_{H^0}$, respectively.}
\label{fig:DD}
\end{figure}

At first glance, it might seem then that the model considered is basically a one-component DM model, given the almost negligible contribution of $H^0$ to the total DM relic density. Nevertheless, the compensation between the very large $Z$ coupling of $H^0$ and its very small relic density makes the fraction of scattering events associated with $H^0$ to be small enough to satisfy the limits on $\mathcal{N}_\text{events}$ and at the same time large enough for sizable signals to be possible (see Fig. \ref{fig:DD}). In fact, $\mathcal{N}^{H^0}_\text{events}$ can be equal or even greater than the singlet contribution $\mathcal{N}^{\phi}_{events}$. The most conservative prospects for the future LZ commissioning shows, for instance, that a significant portion of the models that would be excluded (black points not overlapping with the red ones in Fig.~\ref{fig:DD}) are characterized by a greater proportion ($\gtrsim 70\%$) of expected events associated with $H^0$. In this way, we see that either or even both DM particles are susceptible to detection in future searches.  

To understand these results, let us recall that the strong constraints on Higgs- and $Z$-portal models like the one considered arise due to the tension between efficient DM annihilation in the early universe, which requires large $\lambda_8$ and $\lambda_L$ couplings, and the bounds on DM-nuclei scattering cross sections, which demand small values of these couplings. In this model, however, this tension is relaxed due to the presence of the $\lambda_7$ interaction, which induces semi-annihilation processes during the freeze-out. These processes contribute to reduce the relic abundances of $\phi$ and $H^0$ without the need for large Higgs couplings or large DM masses.  Such a relaxion allows us then to conclude that the effects introduced by the $Z_6$ symmetry enable a scenario capable of satisfying, for DM masses at the EW scale, the stringent experimental restrictions. 
Contrary to what is usually assumed in the literature, a scalar DM model in which the $Z$ portal interactions are relevant (and even dominant) can be perfectly viable, thanks to semi-annihilations which naturally arise in a multi-component DM framework. 

\begin{figure}[t]
\centering
\includegraphics[scale=0.45]{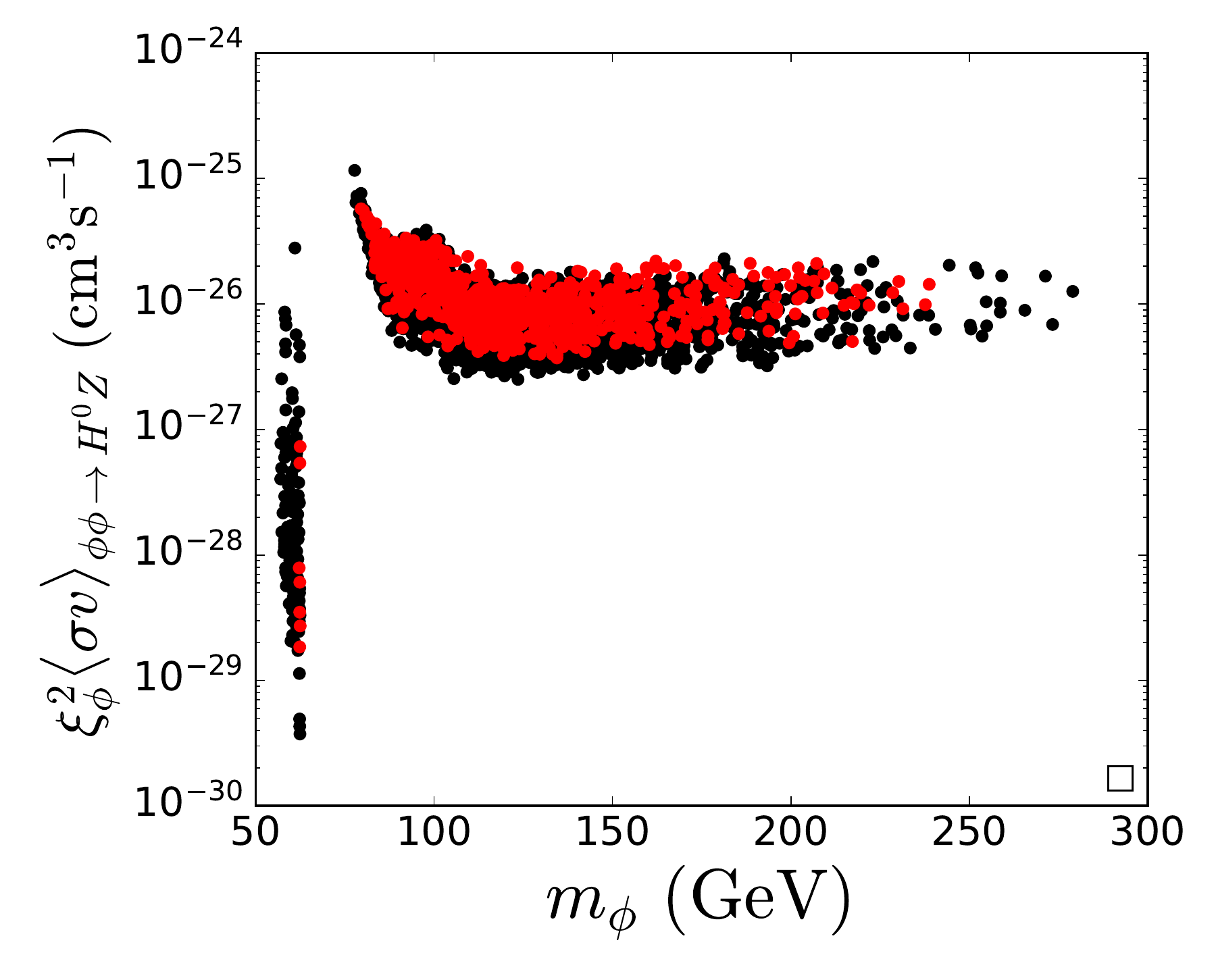}
\includegraphics[scale=0.45]{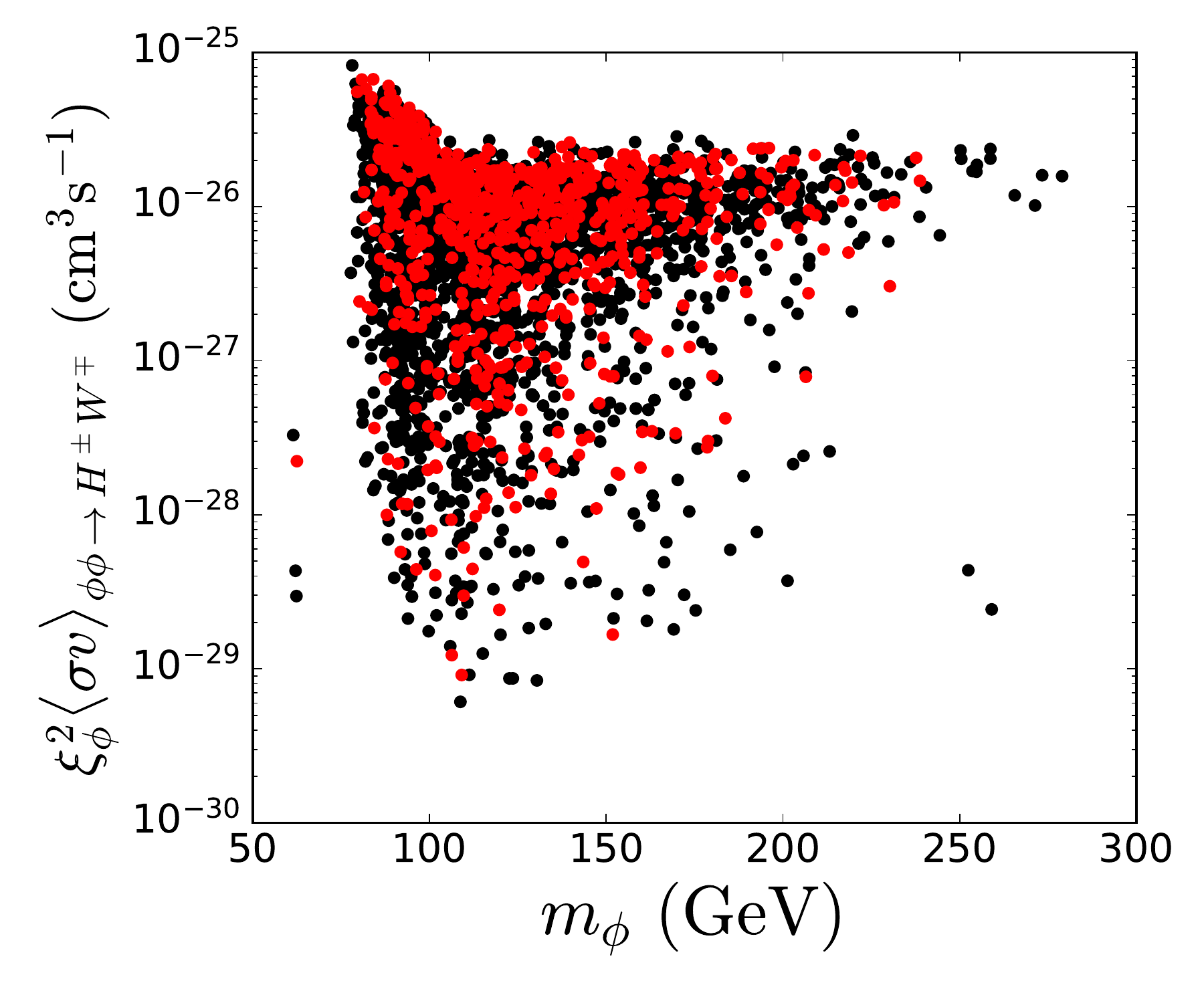}
\caption{Most relevant semi-annihilation processes for ID searches. Scaled cross sections are shown for $\phi\phi\rightarrow H^0Z$ (left panel) and $\phi\phi\rightarrow H^\pm W^\mp$ (right panel). }
\label{fig:ID}
\end{figure}

Regarding DM indirect searches, the most relevant annihilation signals originate from the semi-annihilation channels $\phi \phi \rightarrow H^0 Z$ and $\phi \phi \rightarrow H^\pm W^\mp$, with rates as large as $10^{-25}$ cm$^3$/s. 
As there are no dedicated studies addressing ID through semi-annihilations in multi-component DM scenarios, we just display the corresponding rates scaled by $\xi_\phi^2$ in Fig.~\ref{fig:ID}. 
As a result, it remains unclear whether forthcoming observations will explore the viable parameter space of this model. This uncertainty underscores the need for a focused study to reach definitive conclusions.

A final comment is in order with respect to scenarios featuring larger $SU(2)_L$ multiplets with non-vanishing hypercharge. 
Firstly, let us consider an $SU(2)_L$ multiplet $H_n$ of dimension $n>2$, as opposed to a doublet. Should $\phi$ remain a singlet, an important observation emerges: the operator $\mathcal{O}_n=H_n^{(\dagger)}H_1\, \phi^2$ is no longer invariant under $SU(2)_L$. Therefore, the two-component scalar scenario having a singlet and a multiplet larger than a doublet does not present semi-annihilation processes of the type $\mathcal{O}_n$. 
Next, we go through the case where the singlet field is promoted to a $SU(2)_L$ multiplet $\phi_m$ of dimension $m>1$ with the same hypercharge of the singlet $\phi$, that is $Y(\phi_m)=0$, which in turn demands that $m$ is odd~\cite{Cirelli:2005uq, Hambye:2009pw}. The corresponding operator associated with semi-annihilations takes the form $\mathcal{O}_{nm}=H^\dagger_n\, H_1\, \phi^2_m$, and is gauge invariant for $n$ even and $Y(H_n)=1$. These $(m,n)$ scenarios resemble to some extent the singlet-doublet two-component model when the neutral component of $\phi_m$, $\phi_m^0$, becomes the lightest candidate, and the second candidate being the neutral particle with tree-level $Z$ interactions associated with $H_n$, $H^0_n$. However, thermal relics belonging to higher multiplets possess very large masses, usually in the range of several TeV~\cite{Cirelli:2007xd, Hambye:2009pw}, indicating that the exponential suppression on $\Omega_{H^0_n}$ would not be enough efficient to attenuate the DD rates of $H^0_n$. 

\section{Conclusions}\label{sec:Conclusions}
In this work, we demonstrated that in multi-component scenarios featuring a scalar doublet candidate with unsuppressed vector-like neutral-current gauge interactions with the $Z$ boson, the $Z$-portal can still be open because of the effect of the semi-annihilation processes on the relic abundance on such a candidate. 
For this goal, we enlarged the SM scalar sector with a second Higgs doublet $H_2$ and a complex singlet $\phi$, both with vanishing vacuum expectation value.  A $Z_6$ symmetry prevents the commonly invoked mass splitting of the real and imaginary components of the neutral member of the doublet, $H_0$. As a result, it remains complex, leading to a two-component scalar DM scenario with semi-annihilation processes induced by the interaction term $H^\dagger_2\, H_1\, \phi^2$.

We have shown that these semi-annihilations drastically reduce the relic abundance of $H^0$ by at least seven orders of magnitude relative to that of $\phi$, implying that the $\phi$ component accounts for almost all of the observed DM in the universe. However, 
the smallness of $\Omega_{H^0}$ counteracts the effect of the $Z$-mediated elastic scattering between $H^0$ particles and nuclei,  making signatures from $H^0$ elusive enough to satisfy the most stringent current limits while remaining detectable in ongoing and future DD facilities. Remarkably, these results are obtained for a range of non-degenerate DM masses well below the TeV scale. 

\section*{Acknowledgments}
M.J.R. and O.R. acknowledge the ﬁnancial support given by the  UdeA/CODI Grant 2020-33177. O.R. has received further funding from MinCiencias through the Grant 80740-492-202
O.Z. has been partially supported by Sostenibilidad-UdeA, the UdeA/CODI Grants 2022-52380 and  2023-59130, and Ministerio de Ciencias Grant CD 82315 CT ICETEX 2021-1080.  O.Z. would like to acknowledge support from the ICTP through the Associates Programme (2023-2028).

\bibliographystyle{utphys}
\bibliography{references}

\end{document}